\documentclass[10pt,twocolumn,aps,pra,notitlepage,showpacs,floatfix,amsmath,amssymb;twocolumn]{revtex4-1}
\usepackage[latin9]{inputenc}
\usepackage{amsmath}
\usepackage{graphicx}
\usepackage{wasysym}
\usepackage{esint}

\makeatletter

\newcommand*\LyXThinSpace{\,\hspace{0pt}}

\usepackage{float}

\usepackage{epstopdf}
 \usepackage{setspace}

\makeatother

\begin{document}

\title{A strongly interacting Sarma superfluid near orbital Feshbach resonances}

\author{Peng Zou$^{1}$, Lianyi He$^{2}$, Xia-Ji Liu$^{3}$, and Hui Hu$^{3}$}

\affiliation{$^{1}$College of Physics, Qingdao University, Qingdao 266071, China}

\affiliation{$^{2}$Department of Physics and State Key Laboratory of Low-Dimensional
Quantum Physics, Tsinghua University, Beijing 100084, China}

\affiliation{$^{3}$Centre for Quantum and Optical Science, Swinburne University
of Technology, Melbourne 3122, Australia}

\date{\today}
\begin{abstract}
We investigate the nature of superfluid pairing in a strongly interacting
Fermi gas near orbital Feshbach resonances with spin-population imbalance
in three dimensions, which can be well described by a two-band or
two-channel model. We show that a Sarma superfluid with gapless single-particle
excitations is favored in the closed channel at large imbalance. It
is thermodynamically stable against the formation of an inhomogeneous
Fulde\textendash Ferrell\textendash Larkin\textendash Ovchinnikov
superfluid and features a well-defined Goldstone-Anderson-Bogoliubov
phonon mode and a massive Leggett mode as collective excitations at
low momentum. At large momentum, the Leggett mode disappears and the
phonon mode becomes damped at zero temperature, due to the coupling
to the particle-hole excitations. We discuss possible experimental
observation of a strongly interacting Sarma superfluid with ultracold
alkaline-earth-metal Fermi gases.
\end{abstract}

\pacs{03.75.Ss, 67.85.Lm}
\maketitle

\section{Introduction}

A Sarma phase, named after the pioneering work by Sarma in 1963 \cite{Sarma1963},
is a possible candidate state for a \emph{homogeneous} Fermi superfluid
with pair-breaking population imbalance. Having gapless fermionic
excitations, this state can be conveniently viewed as a phase separation
phase in momentum space: some fermions pair and form a superfluid,
while others occupy in certain regions of momentum space bounded by
gapless Fermi surfaces and remain unpaired. For this reason, the Sarma
phase is also vividly referred to as an interior-gap superfluid \cite{Liu2003,Wu2003}
or a breached-pair superfluid \cite{Forbes2005}. Although the Sarma
phase was predicted more than 50 years ago, its experimental observation
remains elusive, in spite of enormous efforts both experimentally
and theoretically (for recent reviews, see, for example, Refs. \cite{Radzihovsky2010,Chevy2010,Gubbels2013,Kinnunen2017}).
In the original proposal \cite{Sarma1963}, the Sarma phase is a local
maximum solution in the landscape of the grand thermodynamic potential
and suffers from the instability \cite{Wu2003} towards a more stable
phase-separation phase in real space \cite{Bedaque2003} or a spatially
inhomogeneous Fulde\textendash Ferrell\textendash Larkin\textendash Ovchinnikov
(FFLO) superfluid \cite{Fulde1964,Larkin1964}. The realization of
a thermodynamically stable Sarma phase therefore becomes a long-standing
quest \cite{Boettcher2015PLB}.

The recent intensive research interests on the Sarma phase are largely
triggered by the bold proposition by Liu and Wilczek \cite{Liu2003}
and by the rapid experimental progress in ultracold atomic Fermi gases
\cite{Zwierlein2006,Partridge2006,Liao2010}. It has been now realized
that, to cure the instability of the Sarma phase, one needs to carefully
engineer the inter-particle interactions and/or the mass ratio of
the different spin components \cite{Forbes2005,Parish2007,Baarsma2010,Wang2017}.
In the context of two-component spin-1/2 atomic Fermi gases at the
crossover from a Bose-Einstein condensation (BEC) to a Bardeen-Cooper-Schrieffer
(BCS) superfluid \cite{Bloch2008,Giorgini2008,Randeria2014}, the
Sarma phase becomes stable on the BEC side of a BEC-BCS crossover
\cite{Sheehy2006,Sheehy2007,Hu2006PRA}, featuring one gapless Fermi
surface and behaving similar to a Bose-Fermi mixture. A large mass
ratio may greatly enlarge the phase space of the Sarma phase, making
it energetically favorable even at the cusp of the BEC-BCS crossover
\cite{Parish2007,Baarsma2010,Wang2017}, the so-called unitary limit.
In this respect, heteronuclear Fermi-Fermi mixtures of $^{6}$Li-$^{40}$K,
$^{6}$Li-$^{87}$Sr and $^{6}$Li-$^{173}$Yb atoms look very promising,
although there are still some technical issues related to atom loss
and temperature cooling. Interestingly, the Sarma phase may also be
stabilized by considering a multi-band structure. In an early study
\cite{He2009}, one of the present authors showed that in a two-band
Fermi system with four spin components, the inter-band exchange interaction
together with \emph{asymmetric} intra-band interactions can remove
the Sarma instability and the Sarma phase could be the energetically
stable ground state in visible parameter space. It is then natural
to ask, can we realize this kind of two-band proposal with ultracold
atoms?

This possibility may come to true, thanks to the recent innovative
proposal by Zhang, Cheng, Zhai and Zhang \cite{Zhang2015}, named
as orbital Feshbach resonance (OFR), which has been confirmed soon
experimentally \cite{Pagano2015,Hofer2015}. In Fermi gases of alkali-earth
metal atoms (i.e., Sr) or alkali-earth metal like atoms (i.e., Yb),
the long-lived meta-stable orbital (i.e., electronic) state $^{3}P_{0}$
(denoted as $\left|e\sigma\right\rangle $ where $\sigma=\uparrow,\downarrow$
stands for the two internal nuclear spin states) can be selected,
together with the ground orbital state $^{1}S_{0}$ ($\left|g\sigma\right\rangle $).
This forms an effective four-component Fermi system, in which a pair
of atoms can be well described by using the singlet ($-$) and triplet
($+$) basis in the absence of external Zeeman field \cite{Zhang2015,Pagano2015,Hofer2015,He2016},
\begin{equation}
\left|\pm\right\rangle =\frac{1}{2}\left(\left|ge\right\rangle \pm\left|eg\right\rangle \right)\otimes\left(\left|\uparrow\downarrow\right\rangle \mp\left|\downarrow\uparrow\right\rangle \right),
\end{equation}
or by using the two-channel basis in the presence of Zeeman field
\cite{Zhang2015,Pagano2015,Hofer2015,He2016},
\begin{eqnarray}
\left|o\right\rangle  & = & \frac{1}{\sqrt{2}}\left(\left|-\right\rangle +\left|+\right\rangle \right)=\frac{1}{\sqrt{2}}\left(\left|g\uparrow,e\downarrow\right\rangle -\left|e\downarrow,g\uparrow\right\rangle \right),\label{eq:open_channel_basis}\\
\left|c\right\rangle  & = & \frac{1}{\sqrt{2}}\left(\left|-\right\rangle -\left|+\right\rangle \right)=\frac{1}{\sqrt{2}}\left(\left|g\downarrow,e\uparrow\right\rangle -\left|e\uparrow,g\downarrow\right\rangle \right),\label{eq: closed_channel_basis}
\end{eqnarray}
where $o$ and $c$ stand for the open and closed channels, respectively.
The inter-particle interactions are characterized by two underlying
$s$-wave scattering lengths, the singlet scattering length $a_{s-}$
and the triplet scattering length $a_{s+}$, whose magnitude depend
on the atomic species. For $^{173}$Yb atoms, the triplet scattering
length is very large $a_{s+}\simeq1900a_{0}$ \cite{Pagano2015,Hofer2015},
where $a_{0}$ is the Bohr radius. It drives the system into the strongly
interacting regime and also allows one to tune the effective inter-particle
interactions in the open channel via the external Zeeman field \cite{Zhang2015},
which thereby realizes the OFR. It turns out that a Fermi gas near
OFR can be microscopically described by the two-band theory with a
specific form of the interaction Hamiltonian \cite{Zhang2015,He2016,He2015TwoBandTheory}.
To date, a number of many-body effects of a balanced Fermi gas near
OFR have been addressed, including the internal Josephson effect \cite{Iskin2016},
critical temperature \cite{Xu2016}, stability \cite{He2016,Iskin2016},
equations of state \cite{He2016}, collective modes \cite{He2016,Zhang2017},
superfluid properties in a harmonic trap \cite{Iskin2017} and most
recently the closed-channel contributions \cite{Mondal2017}. The
polaron physics in the limit of extreme spin-population imbalance
has also been considered \cite{Chen2016,Xu2017,Chen2018}.

In this work, we would like to confirm the existence of an energetically
stable Sarma superfluid near OFR in three dimensions. This is by no
means obvious from the previous work \cite{He2009}, since the intra-band
interaction potentials are now symmetric within OFR. We also explicitly
explore the stability of the Sarma superfluid against an inhomogeneous
FFLO superfluid. Moreover, for a possible experimental observation,
we consider the zero-temperature collective modes of the Sarma superfluid
and show the existence of a well-defined Leggett mode \cite{Leggett1966}
at low momentum and also a damped Goldstone-Anderson-Bogoliubov phonon
mode at large momentum (due to the coupling to the particle-hole excitations
near the gapless Fermi surface), both of which can be experimentally
probed by using Bragg spectroscopy \cite{Lingham2014}.

The rest of the paper is organized as follows. In the next section
(Sec. II), we introduce the microscopic model of a three-dimensional
strongly interacting Fermi gas near OFR with spin-population imbalance
and outline the mean-field approach to treat different candidate phases
for imbalanced superfluidity, including the Sarma phase and the FFLO
phase. In Sec. III, we examine different imbalanced superfluid states
and show that the Sarma phase is energetically favorable in certain
parameter space. We determine the phase diagram as a function of the
chemical potential difference, for the two cases with a fixed chemical
potential and with a fixed total number of atoms. In Sec. IV, we consider
the Gaussian pair fluctuations on top of the mean-field saddle-point
solution and calculate the Green function of Cooper pairs, from which
we determine the collective modes of either the gapless phonon mode
or the massive Leggett mode. The collective modes of a BCS superfluid
and a Sarma superfluid are explored in a comparative way. The discussions
of the two-particle continuum and the particle-hole continuum of a
Sarma superfluid are given in Appendix A and Appendix B, respectively.
Finally, in Sec. V we draw our conclusions. 

\section{Model Hamiltonian}

We start by an appropriate description of the interaction Hamiltonian
for a Fermi gas near OFR. In the singlet and triplet basis, the interaction
potentials between a pair of atoms can be well approximated by using
pseudo-potentials \cite{Zhang2015,He2016}, 
\begin{equation}
V_{\pm}(r)\simeq\frac{4\pi\hbar^{2}a_{s\pm}}{M}\delta(\mathbf{r})\frac{\partial}{\partial r}\left(r\cdot\right),
\end{equation}
where $M$ is the mass of fermionic atoms. As we use the external
Zeeman field as a control knot, it is convenient to use the two-channel
description, in which, following the basis transform of Eq. (\ref{eq:open_channel_basis})
and Eq. (\ref{eq: closed_channel_basis}), the interaction potentials
become
\begin{eqnarray}
V_{oo}(r) & = & V_{cc}(r)=\frac{V_{-}+V_{+}}{2}=\frac{4\pi\hbar^{2}a_{s0}}{M}\delta(\mathbf{r})\frac{\partial}{\partial r}\left(r\cdot\right),\\
V_{oc}(r) & = & V_{co}\left(r\right)=\frac{V_{-}-V_{+}}{2}=\frac{4\pi\hbar^{2}a_{s1}}{M}\delta(\mathbf{r})\frac{\partial}{\partial r}\left(r\cdot\right).
\end{eqnarray}
The two scattering lengths $a_{s0}$ and $a_{s1}$ are given by, $a_{s0}=(a_{s-}+a_{s+})/2$
and $a_{s1}=(a_{s-}-a_{s+})/2$. The above interaction potentials
can be further replaced with contact potentials, $V_{nm}(r)=V_{nm}\delta(\mathbf{r})$,
with the bare interaction strengths $V_{nm}$ ($n,m=o,c$) to be renormalized
using the two scattering lengths $a_{s0}$ and $a_{s1}$, following
the standard renormalization procedure \cite{He2016}:
\begin{equation}
\left(\begin{array}{cc}
V_{oo} & V_{oc}\\
V_{co} & V_{cc}
\end{array}\right)^{-1}=\frac{M}{4\pi\hbar^{2}}\left(\begin{array}{cc}
a_{0} & a_{1}\\
a_{1} & a_{0}
\end{array}\right)^{-1}-\sum_{\mathbf{k}}\frac{M}{\hbar^{2}\mathbf{k}^{2}}.
\end{equation}
It is then straightforward to write down the interaction Hamiltonian
\cite{He2016},
\begin{equation}
\mathcal{H}_{\textrm{int}}=\sum_{nm}\int d\mathbf{r}V_{nm}\varphi_{n}^{\dagger}\left(\mathbf{r}\right)\varphi_{m}\left(\mathbf{r}\right),
\end{equation}
where $\varphi_{n}(\mathbf{r})\equiv\psi_{n2}(\mathbf{r})\psi_{n1}(\mathbf{r})$
is the field operator of annihilating a pair of atoms in the channel
$n$. For clarity, we use the subscript $i=1,2$ to denote the two
internal degrees of freedom in each channel, instead of using the
spin index $\sigma$. 

The single-particle Hamiltonian in three dimensional free space takes
the standard form \cite{Zhang2015,He2016},
\begin{equation}
\mathcal{H}_{0}=\sum_{ni}\int d\mathbf{r}\psi_{ni}^{\dagger}\left(\mathbf{r}\right)\left(-\frac{\hbar^{2}\nabla^{2}}{2M}-\mu_{ni}\right)\psi_{ni}\left(\mathbf{r}\right),
\end{equation}
where for the channel $n=(o,c)$, we assume $\mu_{n1}=\mu_{n}+\delta\mu_{n}$
and $\mu_{n2}=\mu_{n}-\delta\mu_{n}$. In the presence of a Zeeman
field, a pair of atoms in the open and closed channels has different
Zeeman energy with a difference, $\delta(B)=(g_{g}m_{\downarrow}+g_{e}m_{\uparrow})\mu_{B}B-(g_{g}m_{\uparrow}+g_{e}m_{\downarrow})\mu_{B}B$,
arising from the difference in their magnetic momentum (see Eq. (\ref{eq:open_channel_basis})
and Eq. (\ref{eq: closed_channel_basis})). As a result, we may define
the effective chemical potentials of the open and closed channels
as, $\mu_{o}=\mu$ and 
\begin{equation}
\mu_{c}=\mu-\frac{\delta(B)}{2}.
\end{equation}
In principle, the chemical potential difference in each channel may
be independently tuned experimentally. Throughout the work, we take
\begin{equation}
\delta\mu_{o}=\delta\mu_{c}\equiv\delta\mu,
\end{equation}
since this simple choice captures the essential physics of our work.
It is worth noting that the choice of which channel is open or closed
is somewhat arbitrary. The system remains the same, if one swaps the
label of open and closed channels and simultaneously changes the sign
of the detuning, i.e., $\delta(B)\rightarrow-\delta(B)$. 

We note also that in the previous work \cite{He2009}, the Sarma phase
was found to be stabilized by asymmetric interaction potentials in
the two channels. In our case, the intra-channel interaction potentials
are symmetric (i.e., $V_{oo}=V_{cc}$). However, a nonzero Zeeman
energy difference $\delta(B)\neq0$ introduces an asymmetry in the
single-particle Hamiltonian of the two channels. According to the
OFR mechanism \cite{Zhang2015}, it actually leads to asymmetric \emph{effective}
interaction potentials in the two channels. In this work, we explicitly
examine that the asymmetric effective interaction potentials also
stabilize the Sarma phase.

\subsection{Functional path-integral approach}

We use a functional path-integral approach to solve the three-dimensional
two-band model Hamiltonian, in which the partition function of the
system can be written as \cite{He2016,He2015TwoBandTheory,SadeMelo1993,Hu2006EPL,Diener2008,He2015FermiGas2D},
\begin{equation}
\mathcal{Z}=\int\left[\mathcal{D}\psi\left(x\right)\right]\left[\mathcal{D}\bar{\psi}\left(x\right)\right]\exp\left(-\mathcal{S}\right),
\end{equation}
with an action 
\begin{equation}
\mathcal{S}=\int dx\sum_{ni}\bar{\psi}_{ni}\partial_{\tau}\psi_{ni}\left(x\right)+\int_{0}^{\beta}d\tau\left(\mathcal{H}_{0}+\mathcal{H}_{\textrm{int}}\right).
\end{equation}
Here we use the short-hand abbreviations $x\equiv(\tau,\mathbf{r})$
and $\int dx=\int_{0}^{\beta}d\tau\int d\mathbf{r}$, where $\tau$
is the imaginary time and $\beta\equiv1/(k_{B}T)$ at the temperature
$T$. Following the standard field theoretical treatment \cite{He2016,He2015TwoBandTheory,SadeMelo1993,Diener2008,He2015FermiGas2D},
we use the Hubbard-Stratonovich transformation to decouple the four-field-operator
interaction terms. This amount to setting the auxiliary pairing fields,
\begin{equation}
\mathbf{\boldsymbol{\Phi}}\left(x\right)\equiv\left[\begin{array}{c}
\Phi_{o}\left(x\right)\\
\Phi_{c}\left(x\right)
\end{array}\right]=\left(\begin{array}{cc}
V_{oo} & V_{oc}\\
V_{co} & V_{cc}
\end{array}\right)\left[\begin{array}{c}
\varphi_{o}\left(x\right)\\
\varphi_{c}\left(x\right)
\end{array}\right].
\end{equation}
By integrating out the fermionic degrees of freedom, the partition
function of the system can be rewritten as,
\begin{equation}
\mathcal{Z}=\int\left[\mathcal{D}\boldsymbol{\Phi}\left(x\right)\right]\left[\mathcal{D}\bar{\boldsymbol{\Phi}}\left(x\right)\right]\exp\left(-\mathcal{S}_{\textrm{eff}}\right),
\end{equation}
where the effective action $\mathcal{S}_{\textrm{eff}}$ takes the
form 
\begin{equation}
\mathcal{S}_{\textrm{eff}}=-\int dx\bar{\boldsymbol{\Phi}}\left(\begin{array}{cc}
V_{oo} & V_{oc}\\
V_{co} & V_{cc}
\end{array}\right)^{-1}\boldsymbol{\Phi}-\sum_{n=o,c}\textrm{Tr}\ln\left[-\mathbf{G}_{n}^{-1}\right],
\end{equation}
and the inverse fermionic Green functions are given by
\begin{equation}
\mathbf{G}_{n}^{-1}=\left[\begin{array}{cc}
-\partial_{\tau}+\frac{\hbar^{2}\nabla^{2}}{2M}+\mu_{n1} & \Phi_{n}\left(x\right)\\
\bar{\Phi}_{n}\left(x\right) & -\partial_{\tau}-\frac{\hbar^{2}\nabla^{2}}{2M}-\mu_{n2}
\end{array}\right]\delta\left(x-x'\right).
\end{equation}
In the superfluid phase, the auxiliary pairing fields have nonzero
expectation values. We thus write 
\begin{equation}
\Phi_{n}\left(x\right)=\Delta_{n}\left(\mathbf{r}\right)+\phi_{n}\left(x\right),
\end{equation}
where $\Delta_{o}\left(\mathbf{r}\right)$ and $\Delta_{c}\left(\mathbf{r}\right)$
play the role of the order parameters of the superfluid, and expand
the effective action around the order parameters \cite{He2016,He2015TwoBandTheory,SadeMelo1993,Hu2006EPL,Diener2008},
\begin{equation}
\mathcal{S}_{\textrm{eff}}=\mathcal{S}_{\textrm{MF}}+\mathcal{S}_{\textrm{GF}}\left[\phi_{n},\bar{\phi}_{n}\right]+\cdots.
\end{equation}
In the next subsection (Sec. IIB), we consider the mean-field part
$\mathcal{S}_{\textrm{MF}}$ with order parameters $\Delta_{o}\left(\mathbf{r}\right)$
and $\Delta_{c}\left(\mathbf{r}\right)$. The Gaussian fluctuation
part $\mathcal{S}_{\textrm{GF}}$, which contains the terms quadratic
in $\phi_{n}$ and $\bar{\phi}_{n}$, and the associated low-energy
collective modes will be considered in Sec. IV. All the contributions
beyond the Gaussian level are neglected. 

\subsection{Mean-field theory}

Quite generally, we take the following order parameters \cite{Hu2006PRA,He2006},
\begin{equation}
\Delta_{n}\left(\mathbf{r}\right)=\Delta_{n}e^{-i\mathbf{Q}_{n}\cdot\mathbf{r}},\label{eq: pairing_parameter}
\end{equation}
where $\mathbf{Q}_{n}$ is the center-of-mass momentum of Cooper pairs
in the channel $n=(o,c)$. The standard BCS superfluid or the Sarma
phase has $Q_{n}=0$, while a nonzero $Q_{n}$ implies the possibility
of a FFLO superfluid. Here, for simplicity we consider only the Fulde-Ferrell
(FF) pairing with a plane-wave-like order parameter \cite{Fulde1964}.
More realistic Larkin-Ovchinnikov (LO) pairing (in a standing wave
form \cite{Larkin1964}) and other complicated FFLO pairing schemes
are also possible \cite{Casalbuoni2004}.

By substituting Eq. (\ref{eq: pairing_parameter}) into the fermionic
Green functions and explicitly evaluate $\textrm{Tr}\ln[-\mathbf{G}_{n}^{-1}]$,
we obtain the mean-field thermodynamic potential $\Omega_{\textrm{MF}}=k_{B}T\mathcal{S}_{\textrm{MF}}$,
\begin{eqnarray}
\Omega_{\textrm{MF}} & = & -\mathbf{\Delta}^{\dagger}\left(\begin{array}{cc}
\lambda_{0} & \lambda_{1}\\
\lambda_{1} & \lambda_{0}
\end{array}\right)\mathbf{\Delta}+\sum_{n\mathbf{k}}\left(\xi_{n\mathbf{k}}-E_{n\mathbf{k}}+\frac{M\Delta_{n}^{2}}{\hbar^{2}\mathbf{k}^{2}}\right)\nonumber \\
 &  & -k_{B}T\sum_{n\mathbf{k};\eta=\pm}\ln\left(1+e^{-E_{n\mathbf{k,\eta}}/k_{B}T}\right),\label{eq: Omega}
\end{eqnarray}
where the pairing parameters $\mathbf{\Delta}\equiv(\Delta_{o},\Delta_{c})^{T}$
and we have set the volume to be unity. After renormalization, the
bare interaction strengths have been replaced with $\lambda_{0}$
and $\lambda_{1}$ related to the two scattering lengths:
\begin{eqnarray}
\lambda_{0} & = & +\frac{M}{4\pi\hbar^{2}}\frac{a_{s0}}{a_{s0}^{2}-a_{s1}^{2}},\\
\lambda_{1} & = & -\frac{M}{4\pi\hbar^{2}}\frac{a_{s1}}{a_{s0}^{2}-a_{s1}^{2}}.
\end{eqnarray}
The single-particle dispersion relations in the two channels are given
by, 
\begin{equation}
E_{n\mathbf{k},\pm}=E_{n\mathbf{k}}\pm\left(\frac{\hbar^{2}}{2m}\mathbf{k}\cdot\mathbf{Q}_{n}-\delta\mu\right).
\end{equation}
where 
\begin{eqnarray}
\xi_{n\mathbf{k}} & \equiv & \frac{\hbar^{2}\mathbf{k}^{2}}{2M}-\left(\mu_{n}-\frac{\hbar^{2}Q_{n}^{2}}{8M}\right),\\
E_{n\mathbf{k}} & \equiv & \sqrt{\xi_{n\mathbf{k}}^{2}+\Delta_{n}^{2}}.
\end{eqnarray}
We note that, since we want to check the instability of the Sarma
phase against the FFLO pairing, it is sufficient to consider the possibility
of FF pairing in one channel only. As we are free to swap the label
of the open and closed channels, for concreteness, let us always assume
$\mathbf{Q}_{c}=0$. The center-of-mass momentum of Cooper pairs in
the open channel is allowed to take $\mathbf{Q}_{o}=0$ (BCS or Sarma
pairing) and $\mathbf{Q}_{o}\neq0$ (Fulde-Ferrell pairing).

By minimizing the mean-field thermodynamic potential with respect
to $\Delta_{o}$ and $\Delta_{c}$, we obtain the gap equations,
\begin{eqnarray}
\Delta_{c} & = & \frac{\Delta_{o}}{\lambda_{1}}F_{o}\left(\Delta_{o},Q_{o}\right),\label{eq: GapEquation1}\\
\Delta_{o} & = & \frac{\Delta_{c}}{\lambda_{1}}F_{c}\left(\Delta_{c},Q_{c}\right),\label{eq: GapEquation2}
\end{eqnarray}
where the functions $F_{n}$ ($n=o,c$) are defined as,
\begin{eqnarray}
F_{n}\left(\Delta_{n},Q_{n}\right) & = & -\lambda_{0}+\sum_{\mathbf{k}}\left(\frac{M}{\hbar^{2}\mathbf{k}^{2}}-\frac{1}{2E_{n\mathbf{k}}}\right)\nonumber \\
 &  & +\sum_{\mathbf{k}}\frac{f\left(E_{n\mathbf{k,+}}\right)+f\left(E_{n\mathbf{k,-}}\right)}{2E_{n\mathbf{k}}},
\end{eqnarray}
and $f(x)\equiv1/(e^{x/k_{B}T}+1)$ is the Fermi distribution function.
For the FF pairing, the center-of-mass momentum $Q_{o}$ in the open
channel should also satisfy the saddle point condition,
\begin{equation}
\frac{\partial\Omega_{\textrm{MF}}}{\partial Q_{o}}=0.
\end{equation}
Furthermore, in the case of a fixed total number density $\rho$,
the chemical potential $\mu$ should be adjusted to fulfill the number
equation,
\begin{equation}
\rho=-\frac{\partial\Omega_{\textrm{MF}}}{\partial\mu}.
\end{equation}

\section{Sarma superfluidity}

Throughout the paper, we measure the wavevector and energy in units
of the Fermi wavevector $k_{F}=(3\pi^{2}\rho)^{1/3}$ and the Fermi
energy $\varepsilon_{F}=\hbar^{2}k_{F}^{2}/(2M)$, respectively. In
all the numerical calculations, we set the temperature $T=0$. As
mentioned earlier, for simplicity we assume $\delta\mu_{o}=\delta\mu_{c}\equiv\delta\mu$.
For the interaction strengths, we always take the intra-channel parameter
$1/(k_{F}a_{s0})=-0.5$ and consider two cases of inter-channel coupling:
a \emph{strong} coupling with $1/(k_{F}a_{s1})=-0.05$ and a \emph{weak}
coupling with $1/(k_{F}a_{s1})=-5.0$.

We remark that for a strongly interacting Fermi gas of $^{173}$Yb
atoms near OFR, the typical values for the two coupling strengths
(with density $\rho\simeq5\times10^{13}$ atoms/cm$^{3}$) are $1/(k_{F}a_{s0})\simeq+1.58$
and $1/(k_{F}a_{s0})\simeq-1.95$, respectively \cite{Zhang2015,Pagano2015,Hofer2015,He2016}.
Unfortunately, for this set of interaction parameters, the interesting
many-body physics occurs in an out-of-phase solution of the two pair
potentials (that is, the two order parameters $\Delta_{o}$ and $\Delta_{c}$
have opposite sign), which is metastable only \cite{He2016,Iskin2016}.
In this work, we have tuned the interaction parameters in such a way
that the out-of-phase solution is the absolute many-body ground state.

\begin{figure}
\begin{centering}
\includegraphics[width=0.48\textwidth]{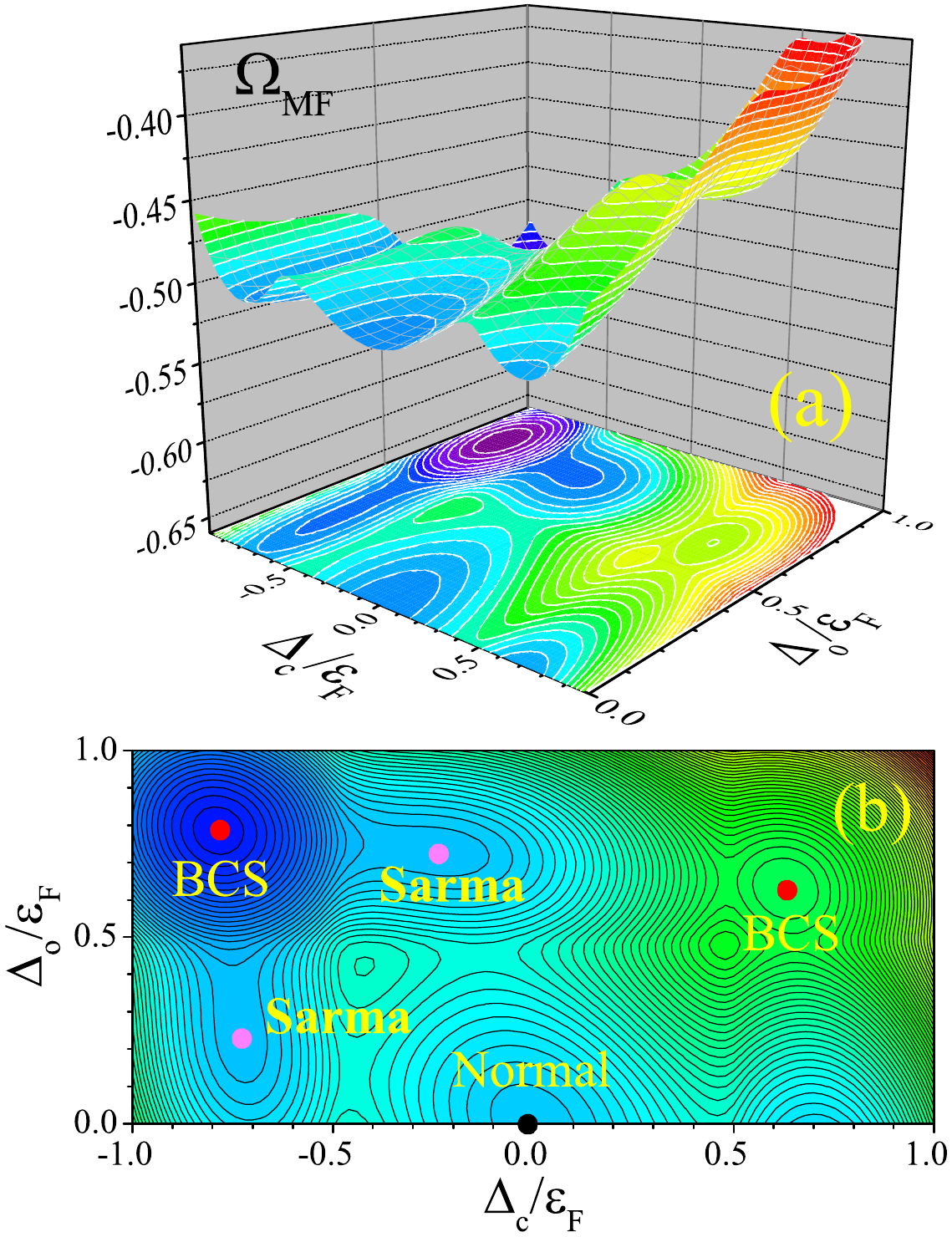} 
\par\end{centering}
\caption{(color online). The landscape of thermodynamic potential (in units
of $N\varepsilon_{F}$) in the $\Delta_{c}-\Delta_{o}$ plane, in
a three-dimensional plot (a) or in a contour plot (b). Here, we take
$1/k_{F}a_{s0}=-0.5$ and $1/k_{F}a_{s1}=-0.05$, $\mu=0.6\varepsilon_{F}$,
$\delta(B)=0$ and $\delta\mu=0.5\varepsilon_{F}$. In the contour
plot, the different competing phases corresponding to the local minima
are indicated. The two channels are symmetric due to zero detuning.
In (b), the in-phase BCS phase, where the two order parameters have
the same sign, is an excited state with an energy much higher than
other local minima. We note that, the FFLO pairing is not allowed
with this set of interaction parameters (i.e., too strong inter-band
coupling). \label{fig1}}
\end{figure}

\begin{figure}
\centering{}\includegraphics[width=0.48\textwidth]{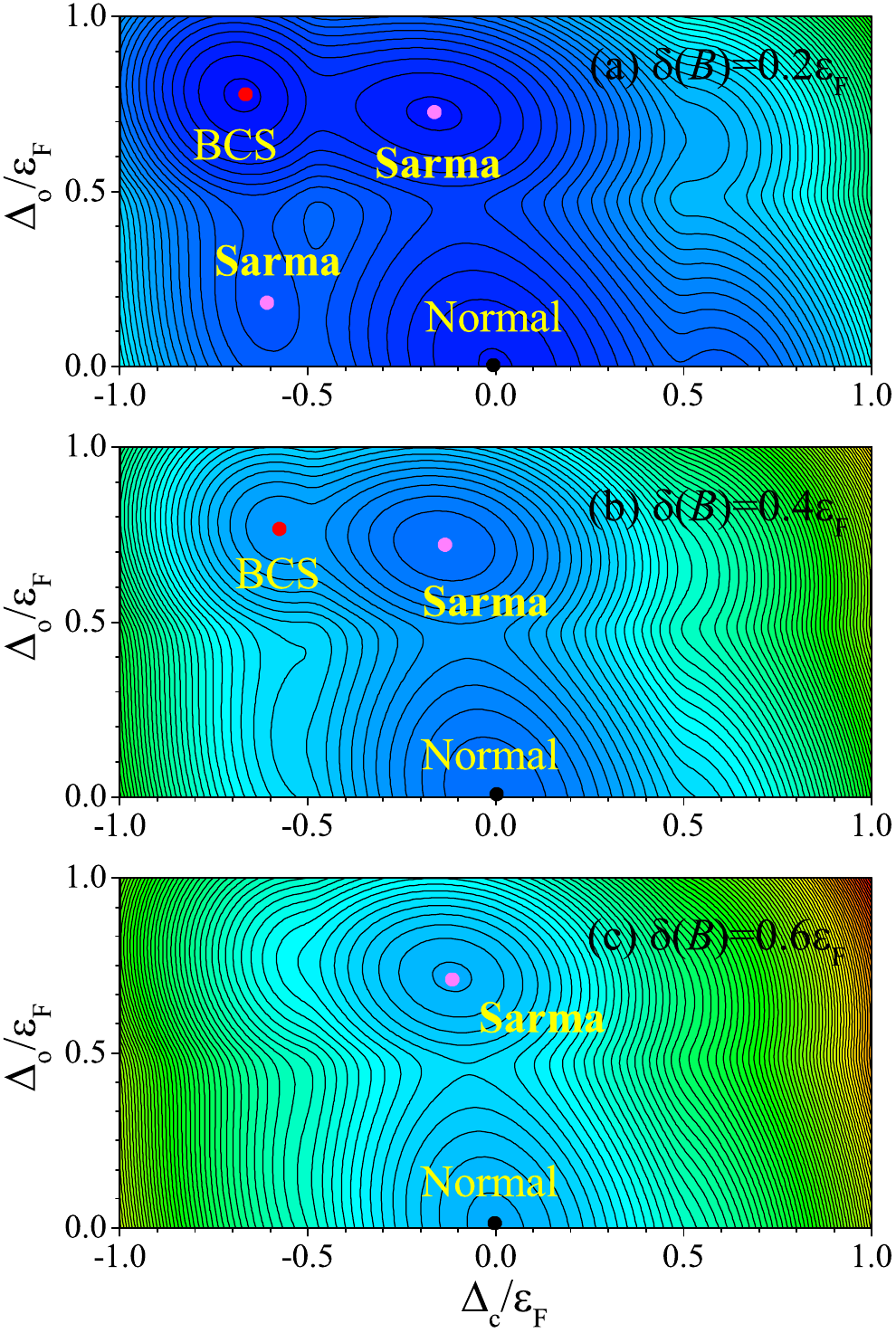}\caption{(color online). The contour plots of thermodynamic potential in the
$\Delta_{c}-\Delta_{o}$ plane, with increasing detuning: (a) $\delta(B)=0.2\varepsilon_{F}$,
(b) $\delta(B)=0.4\varepsilon_{F}$, and (c) $\delta(B)=0.6\varepsilon_{F}$.
The color shows the magnitude of the thermodynamic potential $\Omega_{\textrm{MF}}$,
from blue (small $\Omega_{\textrm{MF}}$) to red (large $\Omega_{\textrm{MF}}$).
Various stable phases (i.e., local minima) are indicated. As the detuning
increases, the pairing type in the closed channel changes from the
BCS to the Sarma pairing. Here, we take $1/k_{F}a_{s0}=-0.5$ and
$1/k_{F}a_{s1}=-0.05$, $\mu=0.6\varepsilon_{F}$ and $\delta\mu=0.5\varepsilon_{F}$.
\label{fig2}}
\end{figure}

\subsection{Sarma pairing at a strong inter-channel coupling}

Figure \ref{fig1} reports various candidate phases of an imbalance
superfluid near OFR with a chemical potential difference $\delta\mu=0.5\varepsilon_{F}$,
in the landscape of the thermodynamic potential. These candidate states
correspond to the local minima in the landscape. Here, we take a strong
inter-channel coupling $1/k_{F}a_{s1}=-0.05$, at which the FFLO pairing
seems to be unfavorable, and work with a grand canonical ensemble,
where the chemical potential is fixed to $\mu=0.6\varepsilon_{F}$.
We also consider a zero detuning $\delta(B)=0$ so that the two channels
are actually symmetric against each other.

It is readily seen that a nonzero chemical potential difference $\delta\mu\neq0$
gives the possibility of Sarma pairing in the open or closed channel,
as indicated in Fig. \ref{fig1}(b). The two Sarma phases, which can
be labelled as {[}BCS{]}$_{o}${[}Sarma{]}$_{c}$ (i.e., $\left|\Delta_{o}\right|>\delta\mu$
and $\left|\Delta_{c}\right|<\delta\mu$) and {[}Sarma{]}$_{o}${[}BCS{]}$_{c}$
($\left|\Delta_{o}\right|<\delta\mu$ and $\left|\Delta_{c}\right|>\delta\mu$)
respectively, should be understood as the same state, due to the equivalence
of the two channels at zero detuning. They are more energetically
favorable than the normal state with vanishing order parameters $\Delta_{o}=\Delta_{c}=0$.
However, they are two local minima only in the landscape of thermodynamic
potential. The global minimum is given by a BCS phase with both order
parameters larger than the chemical potential difference, $\left|\Delta_{o}\right|=\left|\Delta_{c}\right|>\delta\mu$.

The situation dramatically changes when we tune the detuning by switching
on an external Zeeman field. As shown in Fig. \ref{fig2}, with increasing
detuning, the energy of the BCS phase and of one of the Sarma phase
{[}Sarma{]}$_{o}${[}BCS{]}$_{c}$ increases, and both of them disappear
at sufficiently large detuning. In contrast, the Sarma phase {[}BCS{]}$_{o}${[}Sarma{]}$_{c}$
decreases its energy and becomes the global minimum at about $\delta(B)\simeq0.4\varepsilon_{F}$.
Therefore, we find an energetically stable Sarma phase as the absolute
ground state.

Our finding is consistent with the previous observation that the Sarma
phase can be stabilized by introducing an asymmetry between the two
channels or bands \cite{He2009}. However, there is an important difference.
The asymmetry between the two channels in the previous work is caused
by the different intra-channel interaction strengths. In our case,
the intra-channel coupling is always the same. The asymmetry of the
two channels is induced by engineering the single-particle behavior,
i.e., changing the detuning in the closed channel.

\begin{figure}
\centering{}\includegraphics[width=0.48\textwidth]{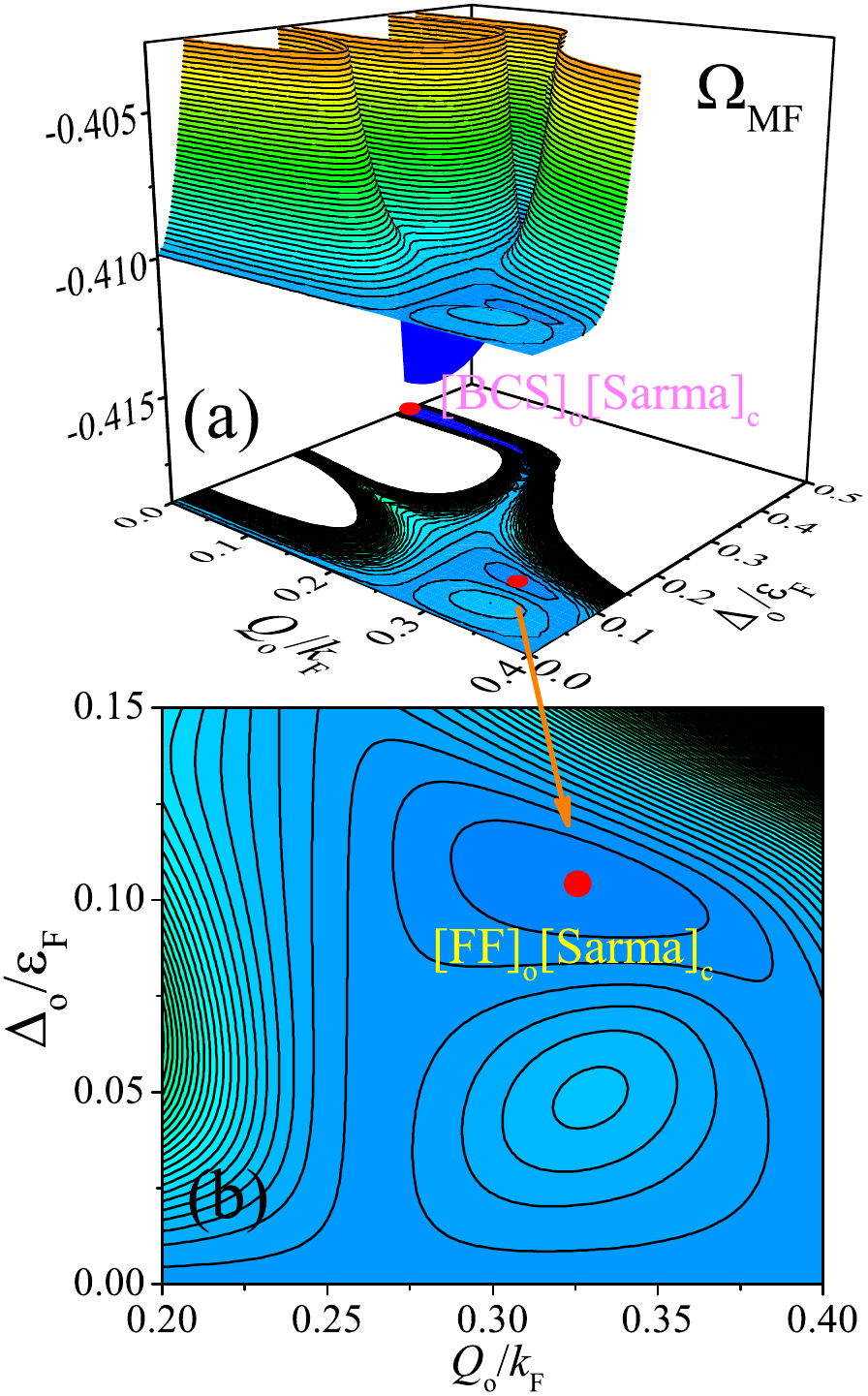}\caption{(color online). The landscape of thermodynamic potential (in units
of $N\varepsilon_{F}$) in the $Q_{o}-\Delta_{o}$ plane, in a three-dimensional
plot (a) or in a zoom-in contour plot to better view the extremely
shallow FF solution (b). Here, we use $1/k_{F}a_{s0}=-0.5$ and $1/k_{F}a_{s1}=-5.0$,
$\mu=0.8\varepsilon_{F}$, $\delta(B)=0.4\varepsilon_{F}$, and $\delta\mu=0.235\varepsilon_{F}$.
\label{fig3}}
\end{figure}

\begin{figure}
\centering{}\includegraphics[width=0.48\textwidth]{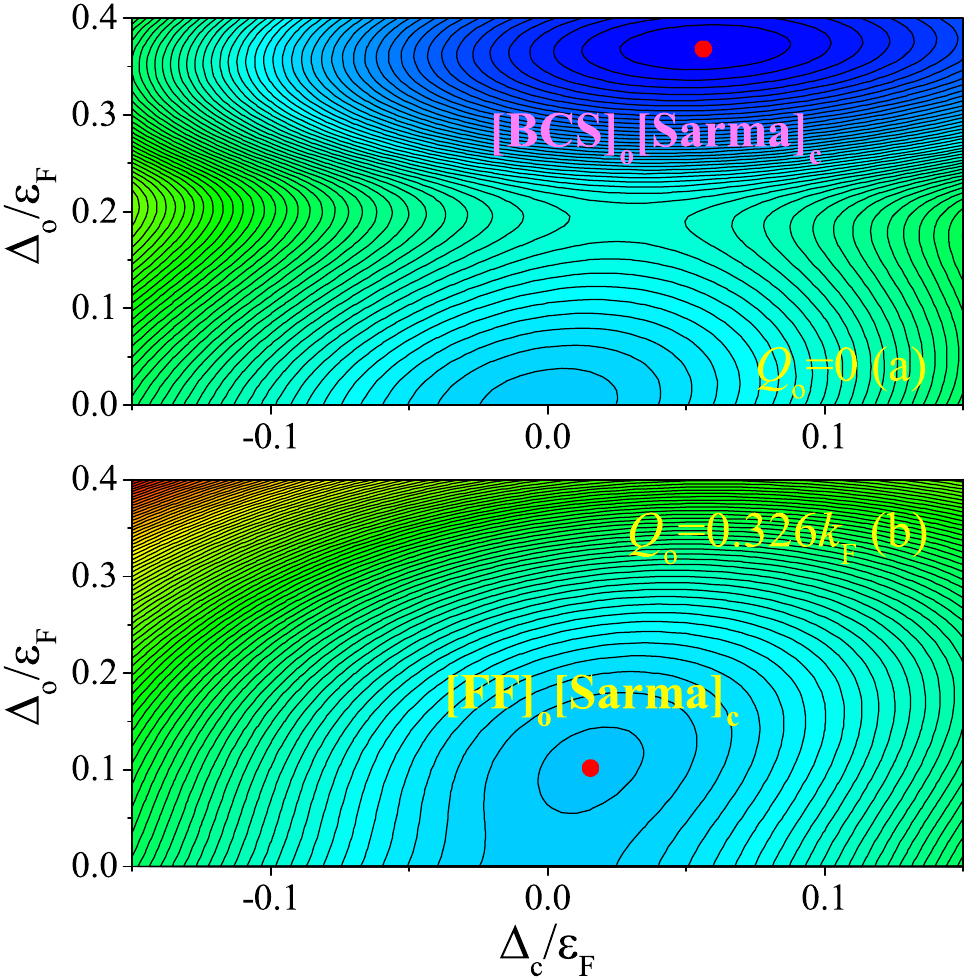}
\caption{(color online). The landscape of thermodynamic potential in the $\Delta_{c}-\Delta_{o}$
plane, with (a) $Q_{o}=0$ and (b) $Q_{o}\simeq0.326k_{F}$. \label{fig4}}
\end{figure}

\subsection{Fulde-Ferrell pairing at a weak inter-channel coupling}

To fully establish the thermodynamic stability of the Sarma phase
in our two-channel model, it is necessary to examine its instability
against the formation of a FFLO superfluid. The latter is often energetically
favorable at large spin-population imbalance in the weak-coupling
limit \cite{Fulde1964,Larkin1964}. To this aim, we choose a weak
inter-channel coupling with $1/k_{F}a_{s1}=-5.0$ and increase slightly
the chemical potential to $\mu=0.8\varepsilon_{F}$. We fix the detuning
to $\delta(B)=0.4\varepsilon_{F}$ and tune the chemical potential
difference $\delta\mu$ to search for the existence of a FF superfluid.

It turns out that the FF pairing in the open channel occurs in a very
narrow interval of the chemical potential difference. In Fig. \ref{fig3},
we present an example at $\delta\mu=0.235\varepsilon_{F}$. The landscape
of thermodynamic potential is shown as functions of the open-channel
order parameter $\Delta_{o}$ and the FF momentum $Q_{o}$. For a
given set of $\Delta_{o}$ and $Q_{o}$, we have used the gap equation
$\Delta_{c}=\Delta_{o}F_{o}(\Delta_{o},Q_{o})/\lambda_{1}$ to determine
the pairing order parameter in the closed channel. It can be seen
from the landscape that a very shallow FF minimum appears at about
$Q_{o}\simeq0.326k_{F}$. To confirm the FF phase is indeed a local
minimum of $\Omega_{\textrm{MF}}(\Delta_{o},\Delta_{c};Q_{o},Q_{c}=0)$,
in Fig. \ref{fig4}(b) we have further checked the contour plot of
thermodynamic potential in the plane of $\Delta_{c}$ and $\Delta_{o}$,
at the optimal momentum of the FF solution $Q_{o}\simeq0.326k_{F}$.
We find that the FF pairing in the open channel is accompanied with
a Sarma pairing in the closed channel, since $\left|\Delta_{c}\right|\ll\delta\mu=0.235\varepsilon_{F}$.
Thus, we denote the FF phase as {[}FF{]}$_{o}${[}Sarma{]}$_{c}$.
For comparison, we also show in Fig. \ref{fig4}(a) the contour plot
of thermodynamic potential at $Q_{o}=0$. It is clear that, at the
chosen parameters, the Sarma phase {[}BCS{]}$_{o}${[}Sarma{]}$_{c}$
has a much lower energy than the FF phase {[}FF{]}$_{o}${[}Sarma{]}$_{c}$. 

\begin{figure}
\centering{}\includegraphics[width=0.48\textwidth]{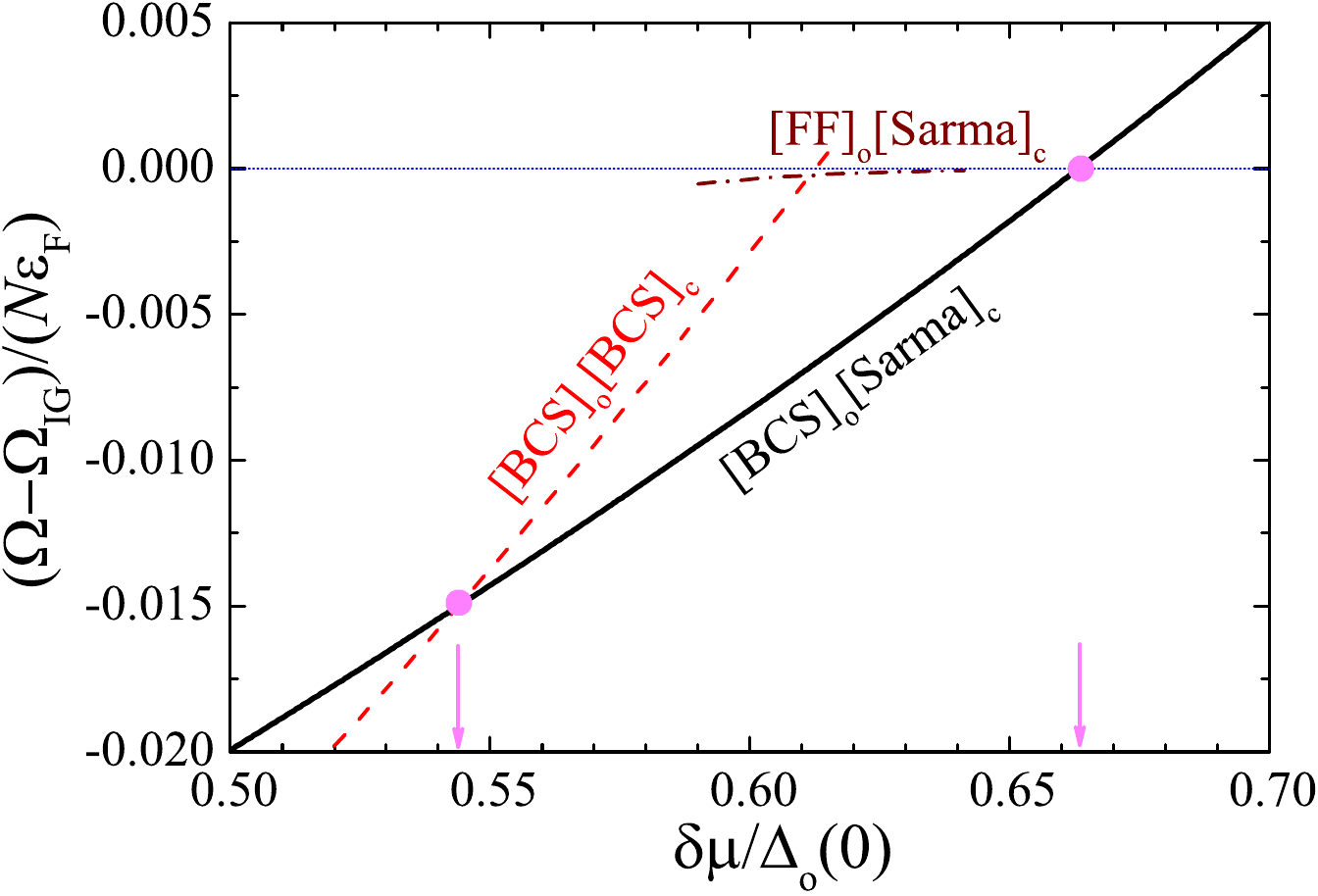}
\caption{(color online). Thermodynamic potentials of different completing phases,
in relative to the ideal Fermi gas result $\Omega_{IG}$ and in units
of $N\varepsilon_{F}$, as a function of the chemical potential difference
$\delta\mu$. The two arrows indicate the positions of the two phase
transitions at $\delta\mu_{c1}\simeq0.543\Delta_{o}(0)$ and $\delta\mu_{c2}\simeq0.662\Delta_{o}(0)$,
respectively. Here, we take $1/k_{F}a_{s0}=-0.5$ and $1/k_{F}a_{s1}=-5.0$,
$\mu=0.8\varepsilon_{F}$, and $\delta(B)=0.4\varepsilon_{F}$. At
$\delta\mu=0$, the pairing gap in the open channel of the low-energy
out-of-phase solution is $\Delta_{o}(0)=0.3898\varepsilon_{F}$. \label{fig5}}
\end{figure}

\subsection{Phase diagram at the weak inter-channel coupling}

By tuning the chemical potential difference $\delta\mu$, we determine
the phase diagram at the given chemical potential $\mu=0.8\varepsilon_{F}$
and at the weak inter-channel coupling, as reported in Fig. \ref{fig5}.
It can be easily seen that, the {[}FF{]}$_{o}${[}Sarma{]}$_{c}$
phase is always not energetically favorable, compared with the {[}BCS{]}$_{o}${[}Sarma{]}$_{c}$
phase. With increasing $\delta\mu$, the imbalanced Fermi gas changes
from the BCS phase ({[}BCS{]}$_{o}${[}BCS{]}$_{c}$) to the {[}BCS{]}$_{o}${[}Sarma{]}$_{c}$
phase, and finally becomes normal. All the transitions are first-order
phase transition.

Therefore, we conclude that a superfluid with the FF pairing form
is not supportive in the two-channel system in three dimensions. Actually,
there is already some indications of this tendency, even if we do
not consider the possibility of the {[}BCS{]}$_{o}${[}Sarma{]}$_{c}$
phase. In the weakly interacting single-channel case, it is well-known
theoretically that a three-dimensional FF superfluid may exist in
the window $0.707\Delta(0)<\delta\mu<0.754\Delta(0)$ \cite{He2006,Casalbuoni2004}.
In our two-channel case, the inter-channel coupling changes the BCS
pairing in the closed channel to the Sarma pairing and also modifies
the window to $0.612\Delta_{o}(0)<\delta\mu<0.641\Delta_{o}(0)$,
which is narrower than the single-channel case.

It is worth noting that for the FFLO superfluid, we may also consider
the LO pairing, which is known to have a lower energy than the FF
pairing. However, in the vicinity of the transition from FFLO to a
normal state, both LO and FF superfluid have very similar energy and
the critical chemical potential difference at the transition will
not change \cite{Casalbuoni2004}, if we use a more accurate LO pairing
order parameter. The only possible change is that, with increasing
$\delta\mu$, we may have a transition from the BCS phase to {[}LO{]}$_{o}${[}Sarma{]}$_{c}$,
and then to {[}BCS{]}$_{o}${[}Sarma{]}$_{c}$. This seems unlikely
to happen.

The situation may qualitatively change if we focus on a low-dimensional
system, where the phase space for FFLO becomes larger \cite{Hu2007,Liu2007,Orso2007,Toniolo2017}.
In that case, intuitively the energy of the {[}FF{]}$_{o}${[}Sarma{]}$_{c}$
phase may become lower than that of the {[}BCS{]}$_{o}${[}Sarma{]}$_{c}$
phase near the superfluid-normal transition. The sequence of phase
transitions is then, {[}BCS{]}$_{o}${[}BCS{]}$_{c}$ $\rightarrow$
{[}BCS{]}$_{o}${[}Sarma{]}$_{c}$ $\rightarrow$ {[}FFLO{]}$_{o}${[}Sarma{]}$_{c}$
$\rightarrow$ normal, with increasing chemical potential difference.
More interestingly, the FFLO pairing may occur in both channels, leading
to the phase {[}FFLO-$Q_{1}${]}$_{o}${[}FFLO-$Q_{2}${]}$_{c}$,
where the FFLO momenta $Q_{1}$ and $Q_{2}$ in the two channels can
be the same or different, depending on the channel coupling. The resultant
rich and complex phase diagram in one-dimension has been recently
explored by Machida and co-workers \cite{Mizushima2013,Takahashi2014},
considering a Pauli-limiting two-band superconductor.

\begin{figure}
\centering{}\includegraphics[width=0.48\textwidth]{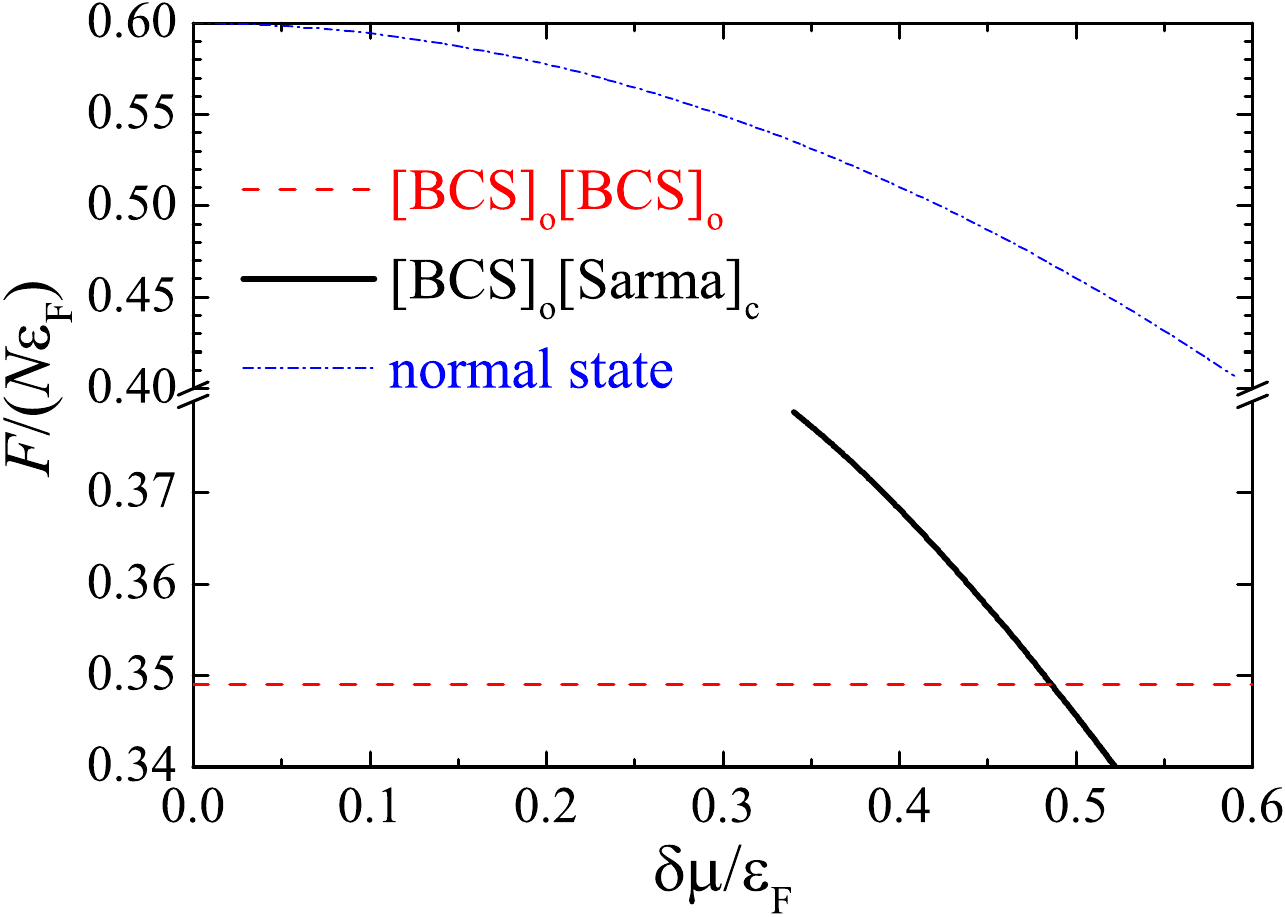}
\caption{(color online). Free energies of different completing phases (in units
of $N\varepsilon_{F}$) as a function of the chemical potential difference
$\delta\mu$, at a fixed total density $\rho=3\pi^{2}k_{F}^{3}$.
The {[}BCS{]}$_{o}${[}Sarma{]}$_{c}$ state becomes favorable above
the threshold $\delta\mu_{c}\simeq0.486\varepsilon_{F}$. Here, we
use $1/k_{F}a_{s0}=-0.5$ and $1/k_{F}a_{s1}=-0.05$, and $\delta(B)=0.4\varepsilon_{F}$.
\label{fig6}}
\end{figure}

\subsection{Phase diagram of free energy with varying chemical potential difference}

We now consider the phase diagram at a fixed total number density
$\rho$ and focus on the case of the strong inter-channel coupling
with $1/k_{F}a_{s1}=-0.05$, in which the FF phase is absent. In Fig.
\ref{fig6}, we report the free energy $F=\Omega+\mu\rho$ of the
three competing phases. In this canonical ensemble for the chosen
parameters, the {[}BCS{]}$_{o}${[}Sarma{]}$_{c}$ appears at $\delta\mu\simeq0.3\varepsilon_{F}$,
becomes energetically favorable at $\delta\mu\simeq0.486\varepsilon_{F}$,
and finally disappears at about $\delta\mu\simeq0.53\varepsilon_{F}$,
with a vanishingly small $\Delta_{c}$. At a large chemical potential
difference $\delta\mu>0.53\varepsilon_{F}$, we thus expect a phase-separation
phase in real space \cite{Bedaque2003}, consisting of both {[}BCS{]}$_{o}${[}BCS{]}$_{c}$
and the normal state.

\section{Collective modes of a Sarma superfluid}

Ideally, a Sarma superfluid may be experimentally detected by measuring
the momentum distribution $\rho(\mathbf{k})$ after time-of-flight
expansion \cite{Yi2006} or by measuring the momentum-resolved radio-frequency
spectroscopy that gives directly the single-particle spectral function
$\mathcal{A}(\mathbf{k},\omega)$ \cite{Stewart2008}. These probes
cannot be universally applied to all fermionic species of ultracold
atoms. For example, for $^{6}$Li atoms near a broad Feshbach resonance,
the strong inter-particle interactions will qualitatively change the
momentum distribution during the early stage of time-of-flight. These
measurements are possible only for atomic species that has a relatively
narrow Feshbach resonance, such as $^{40}$K, in which one can quickly
switch the magnetic field to the non-interacting limit, to avoid the
convert of the interaction energy to the kinetic energy during the
expansion.

A \emph{universal} theme to probe a strongly interacting Sarma superfluid
is provided by Bragg spectroscopy, which measures the density-density
dynamic structure factor and determines low-energy collective excitations
of the system \cite{Lingham2014}. In this section, we discuss the
collective modes of BCS and Sarma phases and show that a Sarma superfluid
in the form of {[}BCS{]}$_{o}${[}Sarma{]}$_{c}$ has some unique
features in its collective excitations, as a consequence of the fact
$\left|\Delta_{c}\right|<\delta\mu$.

Theoretically, the basic information of low-energy collective modes,
such as the mode frequency and the damping rate, can be extracted
from the vertex function, which can be regarded as the Green function
of Cooper pairs in the lowest-order approximation \cite{He2016,Hu2006EPL,Diener2008}.
To obtain the vertex function, we expand the effective action around
the mean-field saddle point and the resultant Gaussian fluctuation
part $\mathcal{S}_{\textrm{GF}}$ is given by,\begin{widetext}
\begin{equation}
\mathcal{S}_{\textrm{GF}}=\frac{1}{2}\sum_{Q}\left[\phi_{o}^{*}\left(Q\right),\phi_{o}\left(-Q\right),\phi_{c}^{*}\left(Q\right),\phi_{c}\left(-Q\right)\right]\left[-\Gamma^{-1}\left(Q\right)\right]\left[\phi_{o}\left(Q\right),\phi_{o}^{*}\left(-Q\right),\phi_{c}\left(Q\right),\phi_{c}^{*}\left(-Q\right)\right]^{T}
\end{equation}
where $Q\equiv(\mathbf{q},i\nu_{l})$ and $i\nu_{l}=i2\pi lk_{B}T$
($l=0,\pm1,\pm2,\cdots$) are the bosonic Matsubara frequencies, and
the inverse vertex function is \cite{He2016}
\begin{equation}
-\Gamma^{-1}\left(\mathbf{q},i\nu_{l}\right)\equiv\left[\begin{array}{cccc}
-\lambda_{0}+M_{11}^{(o)} & M_{12}^{(o)} & -\lambda_{1} & 0\\
M_{21}^{(o)} & -\lambda_{0}+M_{22}^{(o)} & 0 & -\lambda_{1}\\
-\lambda_{1} & 0 & -\lambda_{0}+M_{11}^{(c)} & M_{12}^{(c)}\\
0 & -\lambda_{1} & M_{21}^{(c)} & -\lambda_{0}+M_{22}^{(c)}
\end{array}\right]=-\left[\begin{array}{cc}
\Gamma_{(o)}^{-1}\left(\mathbf{q},i\nu_{l}\right) & \textrm{diag}\{\lambda_{1},\lambda_{1}\}\\
\textrm{diag}\{\lambda_{1},\lambda_{1}\} & \Gamma_{(c)}^{-1}\left(\mathbf{q},i\nu_{l}\right)
\end{array}\right]\label{eq: InverseVertexFunction}
\end{equation}
with the matrix elements at zero temperature ($n=o,c$),
\begin{eqnarray}
M_{11}^{(n)} & = & \sum_{\mathbf{k}}\left[v_{n-}^{2}u_{n+}^{2}\frac{f_{n-}-f_{n+}}{i\nu_{l}+E_{n-}-E_{n+}}+u_{n-}^{2}u_{n+}^{2}\frac{1-f_{n+}}{i\nu_{l}-E_{n-}-E_{n+}}-v_{n-}^{2}v_{n+}^{2}\frac{1-f_{n-}}{i\nu_{l}+E_{n-}+E_{n+}}+\frac{M}{\hbar^{2}\mathbf{k}^{2}}\right],\label{eq: M11}\\
M_{12}^{(n)} & = & \sum_{\mathbf{k}}\left(uv\right)_{n-}\left(uv\right)_{n+}\left[\frac{f_{n-}-f_{n+}}{i\nu_{l}+E_{n-}-E_{n+}}-\frac{1-f_{n+}}{i\nu_{l}-E_{n-}-E_{n+}}+\frac{1-f_{n-}}{i\nu_{l}+E_{n-}+E_{n+}}\right],\label{eq: M12}
\end{eqnarray}
\end{widetext}and $M_{21}^{(n)}(Q)=M_{12}^{(n)}(Q)$, and $M_{22}^{(n)}(Q)=M_{11}^{(n)}(-Q)$.
Here, we have defined some short-hand notations,
\begin{eqnarray}
E_{n\pm} & \equiv & E_{n\mathbf{k}\pm\mathbf{q}/2},\\
f_{n\pm} & \equiv & f\left[E_{n\mathbf{k}\pm\mathbf{q}/2}-\delta\mu\right],\\
u_{n\pm}^{2} & = & \frac{1}{2}\left(1+\frac{\xi_{n\pm}}{E_{n\pm}}\right),\\
v_{n\pm}^{2} & = & \frac{1}{2}\left(1-\frac{\xi_{n\pm}}{E_{n\pm}}\right),\\
\left(uv\right)_{n\pm} & = & \frac{1}{2}\frac{\Delta_{n}}{E_{n\pm}}.
\end{eqnarray}
The low-lying collective excitation spectrum is determined by the
pole of $\Gamma(\mathbf{q},i\nu_{l}\rightarrow\omega+i0^{+})$ after
analytic continuation \cite{He2016,Zhang2017,Matera2017}.

\begin{figure*}
\begin{centering}
\includegraphics[clip,width=0.75\textwidth]{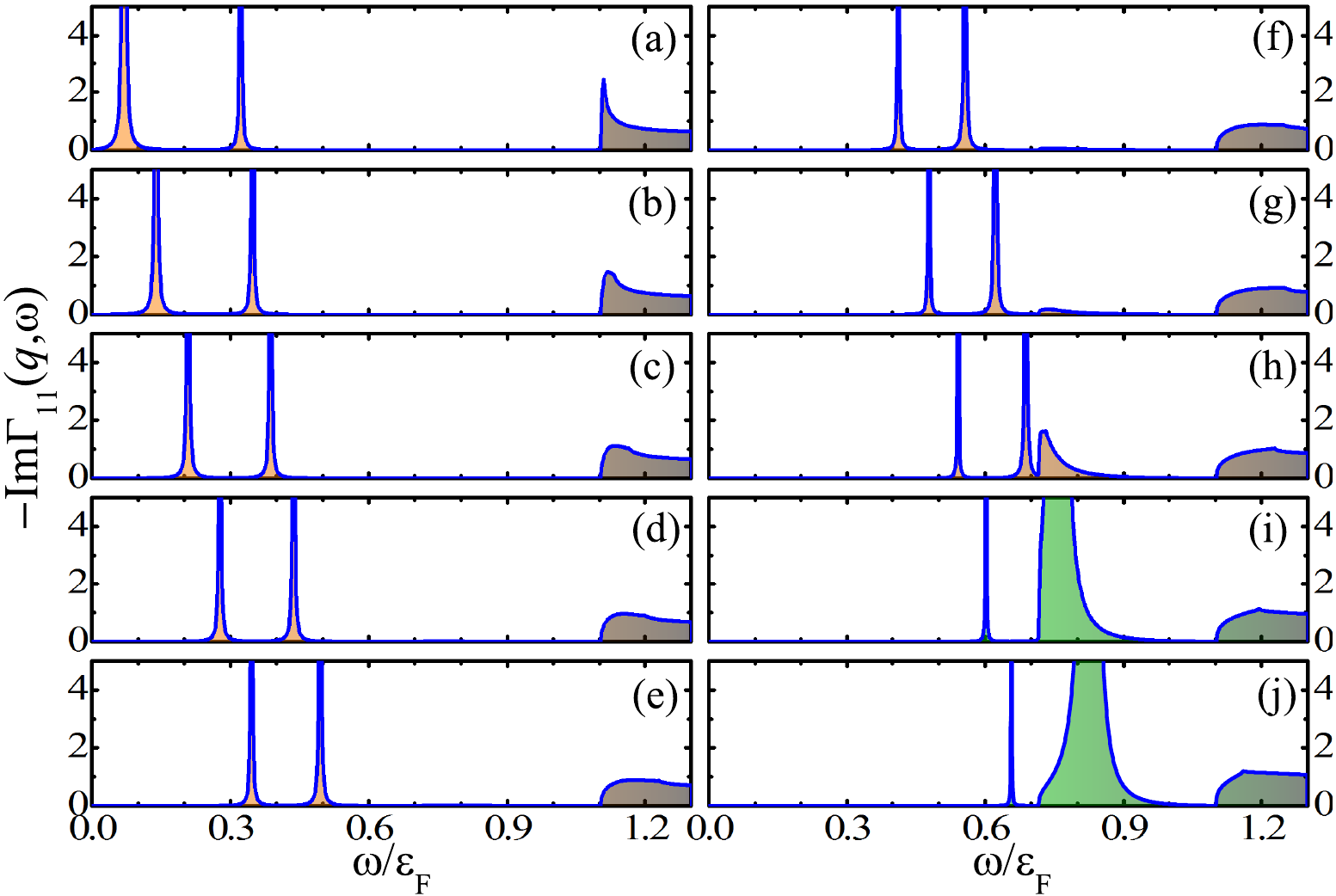} 
\par\end{centering}
\begin{centering}
\includegraphics[clip,width=0.5\textwidth]{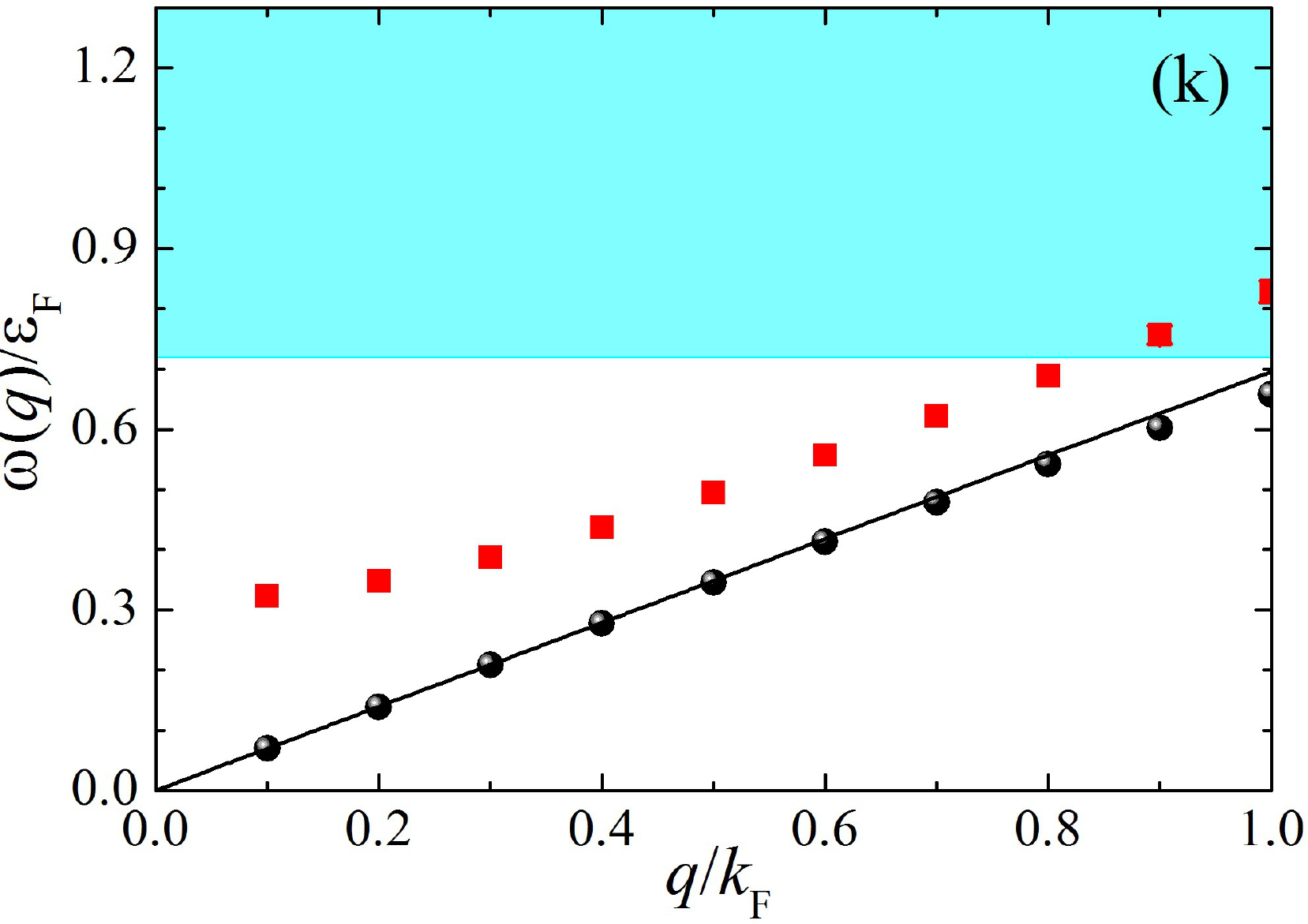} 
\par\end{centering}
\centering{}\caption{(color online). (a)-(j) The spectral function of Cooper pairs $-\textrm{Im}\Gamma_{11}(q,\omega)$
at different transferred momenta, which increase from $0.1k_{F}$
to $1.0k_{F}$ with a step $0.1k_{F}$. Here, we consider a balanced
BCS state with a fixed total density $\rho=3\pi^{2}k_{F}^{3}$ and
$\delta\mu=0$. The detuning is $\delta(B)=0.4\varepsilon_{F}$. (k)
The dispersion relation of the phonon (black circles) and Leggett
modes (red squares). The shaded area in cyan color shows the two-particle
excitation continuum. The error bar in red squares in the two-particle
continuum indicates the damping width of the peak, due to the coupling
to fermionic quasiparticles. The straight line shows $\omega=c_{s}q$,
where $c_{s}\simeq0.348v_{F}$ is the sound velocity. \label{fig7}}
\end{figure*}

It is clear from Eq. (\ref{eq: InverseVertexFunction}) that the total
vertex function is constructed from the two vertex functions in each
channel coupled by the inter-channel coupling matrix $\textrm{diag}\{\lambda_{1},\lambda_{1}\}$,
which is diagonal. In the absence of coupling, it is well-known that
each vertex function supports a gapless Goldstone-Anderson-Bogoliubov
phonon mode, due to phase fluctuations of the pairing order parameter.
As we shall see \cite{He2016,Zhang2017}, with inter-channel coupling,
one gapless mode remains, corresponding to the in-phase phase fluctuations
of the two order parameters. It is ensured by the condition 
\begin{equation}
\det\Gamma^{-1}\left(\mathbf{q}=0,i\nu_{l}=0\right)=0,
\end{equation}
which is exactly equivalent to the gap equations Eq. (\ref{eq: GapEquation1})
and Eq. (\ref{eq: GapEquation2}). The other gapless mode, corresponding
to the out-of-phase phase fluctuations, is lifted to have a finite
energy in the low-wavelength limit. This is the so-called massive
Leggett mode \cite{Leggett1966}, which is not observed with cold-atoms
yet. 

From the expressions Eq. (\ref{eq: M11}) and Eq. (\ref{eq: M12})
of the matrix elements $M_{11}^{(n)}$ and $M_{12}^{(n)}$, one may
easily identify that collective excitations are coupled to the two
types of single-particle excitations: (i) pair-breaking excitations,
in which two Bogoliubov quasi-particles are created or annihilated
with possibility $u_{n-}^{2}u_{n+}^{2}(1-f_{n+})$ or $v_{n-}^{2}v_{n+}^{2}(1-f_{n-})$.
These are given in the second and third terms of the matrix elements;
and (ii) particle-hole excitations, in which one Bogoliubov quasi-particle
is scattered into another quasi-particle state, with possibility proportional
to the number of quasi-particles present, i.e., $f_{n-}-f_{n+}$.
This process is described by the first term of the matrix elements.

\begin{figure*}
\begin{centering}
\includegraphics[clip,width=0.75\textwidth]{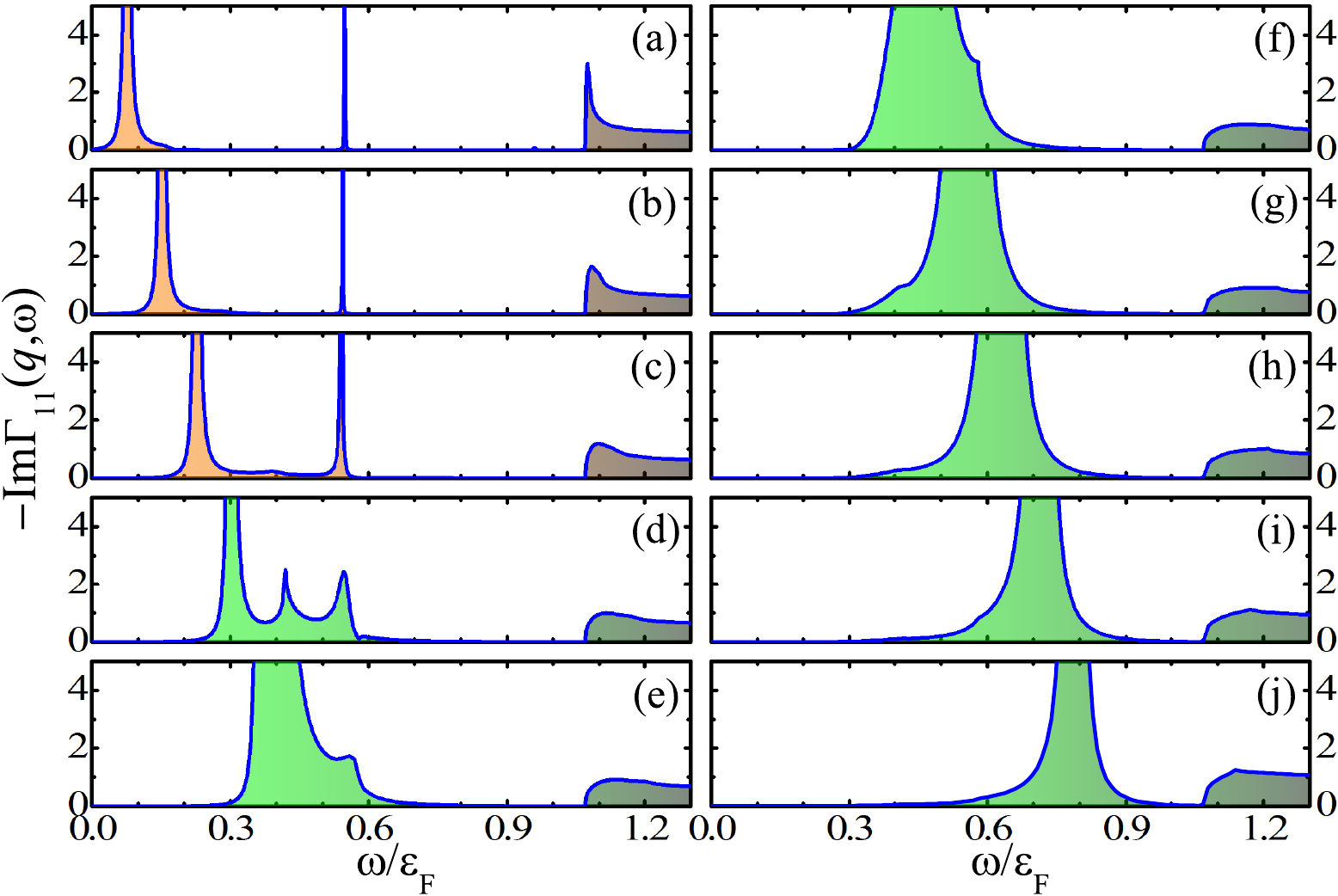} 
\par\end{centering}
\begin{centering}
\includegraphics[clip,width=0.5\textwidth]{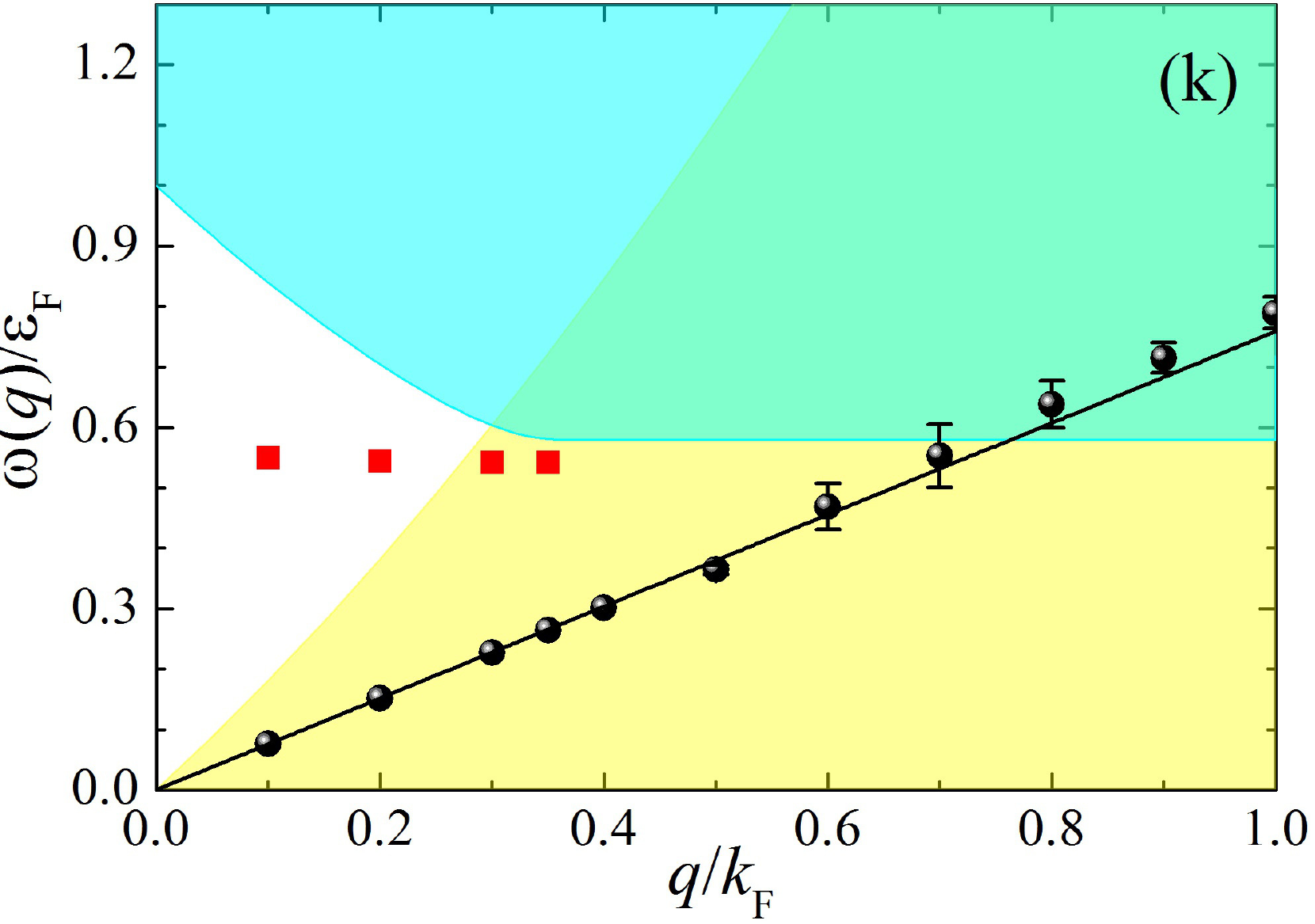} 
\par\end{centering}
\centering{}\caption{(color online). (a)-(j) The spectral function of Cooper pairs $-\textrm{Im}\Gamma_{11}(q,\omega)$
at different transferred momenta, which increase from $0.1k_{F}$
to $1.0k_{F}$ with a step $0.1k_{F}$. Here, we take the parameters
$\delta(B)=0.4\varepsilon_{F}$ and $\delta\mu=0.5\varepsilon_{F}$
and consider the case of a fixed total density $\rho=3\pi^{2}k_{F}^{3}$,
which lead to a Sarma state in the closed channel with $\mu_{c}=\mu-\delta(B)/2\simeq0.25\varepsilon_{F}$
and $\Delta_{c}\simeq-0.08\varepsilon_{F}$. (k) The dispersion relation
of the phonon (black circles) and Leggett modes (red squares). The
shaded areas in cyan and yellow colors correspond to the two-particle
(Eq. (\ref{eq:w2p})) and particle-hole excitation continua (Eq. (\ref{eq:wph})),
respectively. The error bar in symbols indicates the full width at
half maximum of the peak in the spectral function. As the momentum
increases, the Leggett peak shrinks gradually and disappears at $q\sim0.4k_{F}$.
The straight line shows $\omega=c_{s}q$, where $c_{s}\simeq0.380v_{F}$
is the sound velocity. \label{fig8}}
\end{figure*}

\subsection{Collective modes of a BCS superfluid}

In Fig. \ref{fig7}, we report the spectral function of Cooper pairs
- the imaginary part of the 11-component of the vertex function $-\textrm{Im}\Gamma_{11}(q,\omega)$
- of a BCS superfluid, at different transferred momenta from $q=0.1k_{F}$
to $q=k_{F}$, with a step $\Delta q=0.1k_{F}$. Here, we choose the
same interaction parameters as in the phase diagram Fig. \ref{fig6}
and set the chemical potential difference $\delta\mu=0$. The self-consistent
solution of the mean-field equations at a fixed number of atoms $\rho$
leads to, $\mu_{o}\simeq0.43\varepsilon_{F}$, $\mu_{c}\simeq0.23\varepsilon_{F}$,
$\Delta_{o}\simeq0.55\varepsilon_{F}$ and $\Delta_{c}\simeq-0.36\varepsilon_{F}$.

For a BCS superfluid, it is clear that the particle-hole excitations
are absent at zero temperature, as a result of the gapped single-particle
spectrum and that all the fermionic distribution functions $f_{n\pm}$
should vanish identically. Pair-breaking excitations are possible
if the frequency $\omega$ is larger than the two-particle threshold,
which for the channel $n=o,c$ is given by \cite{Combescot2006},
\begin{equation}
\omega_{\textrm{2p}}^{(n)}=\left\{ \begin{array}{cc}
2\sqrt{\left(\frac{\hbar^{2}q^{2}}{8M}-\mu_{n}\right)^{2}+\Delta_{n}^{2}} & \textrm{if }\mu_{n}<\frac{\hbar^{2}q^{2}}{8M}\\
2\left|\Delta_{n}\right| & \textrm{otherwise}
\end{array}\right..
\end{equation}
For small $q$, we thus obtain $\omega_{\textrm{2p}}^{(o)}=2\Delta_{o}\simeq1.10\varepsilon_{F}$
and $\omega_{\textrm{2p}}^{(c)}=2\left|\Delta_{c}\right|\simeq0.72\varepsilon_{F}$.

From the spectral functions at $q\leq0.8k_{F}$, i.e., in Figs. \ref{fig7}(a)-\ref{fig7}(h),
one can clearly identify the gapless Goldstone-Anderson-Bogoliubov
phonon mode and the massive Leggett mode, both of which are undamped,
since they do not touch the two-particle continuum of either channel.
For the cases with $q=0.9k_{F}$ in Fig. \ref{fig7}(i) and with $q=k_{F}$
in Fig. \ref{fig7}(j), the phonon mode remains undamped, while the
Leggett mode has a frequency larger than $\omega_{\textrm{2p}}^{(c)}$
and gets damped due to the coupling to the pair-breaking excitations.
The dispersion relations of the phonon mode and the Leggett mode in
the BCS superfluid are summarized in Fig. \ref{fig7}(k).

\begin{figure}
\centering{}\includegraphics[width=0.48\textwidth]{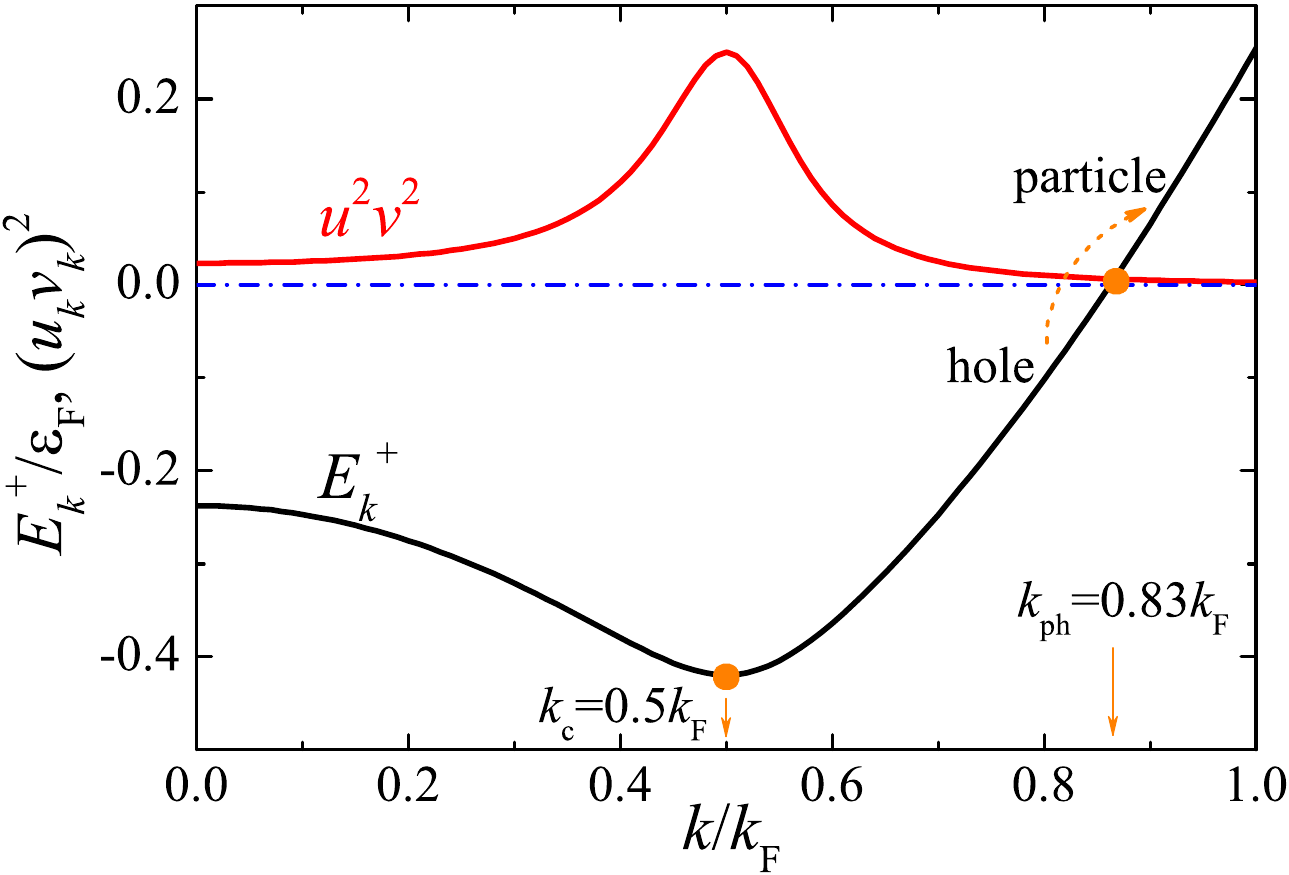}
\caption{(color online). The dispersion relation $E_{c\mathbf{k},+}=\sqrt{\xi_{c\mathbf{k}}^{2}+\Delta_{c}^{2}}-\delta\mu$
of the lower branch of fermionic quasiparticles in the closed channel
(black curve) and the square of the coherence factor $(u_{c\mathbf{k}}v_{c\mathbf{k}})^{2}$
(red curve). Here, we take the parameters $\delta(B)=0.4\varepsilon_{F}$
and $\delta\mu=0.5\varepsilon_{F}$ and consider the case of a fixed
total density $\rho=3\pi^{2}k_{F}^{3}$, which lead to a Sarma state
in the closed channel with $\mu_{c}=\mu-\delta(B)/2\simeq0.25\varepsilon_{F}$
and $\Delta_{c}\simeq-0.08\varepsilon_{F}$. The dispersion relation
has gapless particle-hole excitations near the wavevectors $\left|\mathbf{k}\right|=k_{\mathbf{\textrm{ph}}}\simeq0.83k_{F}$.
The factor $(u_{c\mathbf{k}}v_{c\mathbf{k}})^{2}$ peaks near the
minimum of the dispersion relation ($k_{c}\simeq0.5k_{F}$). \label{fig9}}
\end{figure}

\subsection{Collective modes of a Sarma superfluid}

The collective modes of a Sarma superfluid are quite different, because
of the gapless single-particle excitations. In Fig. \ref{fig8}, we
present the spectral function $-\textrm{Im}\Gamma_{11}(q,\omega)$
of a Sarma superfluid, as the transferred momentum evolves from $q=0.1k_{F}$
to $q=k_{F}$ in step of $\Delta q=0.1k_{F}$. To have a Sarma phase
in the closed channel, we again choose the interaction parameters
as in Fig. \ref{fig6} and take the chemical potential difference
$\delta\mu=0.5\varepsilon_{F}$. The mean-field solution at the fixed
density $\rho$ gives $\mu_{o}\simeq0.45\varepsilon_{F}$, $\mu_{c}\simeq0.25\varepsilon_{F}$,
$\Delta_{o}\simeq0.54\varepsilon$ and $\Delta_{c}\simeq-0.08\varepsilon_{F}$.
For the lower branch of fermionic quasiparticles in the closed channel,
we show its dispersion relation $E_{c\mathbf{k},+}=\sqrt{\xi_{c\mathbf{k}}^{2}+\Delta_{c}^{2}}-\delta\mu$
in Fig. \ref{fig9}. The closed-channel dispersion relation has a
node at $k_{\textrm{ph}}\simeq0.83k_{F}$ and has a minimum at $k_{c}\simeq0.5k_{F}$.

At the smallest $q$ considered, as shown in Fig. \ref{fig8}(a),
we find the anticipated phonon mode and Leggett mode, as in the case
of a BCS superfluid. As $q$ increases, however, the situation becomes
different. While the phonon mode remains well-defined, the massive
Leggett mode gradually loses its weight and finally disappears at
$q\sim0.4k_{F}$. By further increasing $q$, a damped phonon mode
with nonzero damping width is observed, up to the largest transferred
momentum considered in the figure (i.e., $q=k_{F}$). The dispersion
relations of the phonon mode and the Leggett mode in the Sarma superfluid
are summarized in Fig. \ref{fig8}(k).

The vanishing Leggett mode and the damped phonon mode at $q\apprge0.4k_{F}$
may be understood from the gapless single-particle spectrum, which
allows nonzero fermionic distribution functions $f_{n\pm}>0$ and
hence the gapless particle-hole excitations. As discussed in Appendix
B, the collective modes couple to the particle-hole excitations if
their frequency $\omega$ is within the particle-hole continuum, i.e.,
\begin{equation}
0<\omega<\omega_{\textrm{ph}}=\sqrt{\left[\frac{\hbar^{2}(k_{\textrm{ph}}+q)^{2}}{2m}-\mu_{c}\right]+\Delta_{c}^{2}}-\delta\mu.
\end{equation}
This particle-hole continuum has been indicated in Fig. \ref{fig8}(k)
in yellow. We find that the Leggett mode is fragile towards the excitations
of particle-hole pairs. The phonon mode seems to be more robust. In
particular, the damping rate of the phonon mode due to the coupling
to the particle-hole excitations can hardly be noticed at $q\lesssim0.3k_{F}$,
because of the small coherence factors (see Appendix B for more details).

On the other hand, it is somehow surprising to find a well-defined,
undamped Leggett mode at $q\sim0$. Naïvely, one may think that the
two-particle threshold for the closed channel is $\omega_{\textrm{2p}}^{(c)}=2\left|\Delta_{c}\right|\simeq0.16\varepsilon_{F}$.
As the frequency of the Leggett mode is about $\omega\sim0.6\varepsilon_{F}>\omega_{\textrm{2p}}^{(c)}$,
it should be damped by the process of breaking a Cooper pair. This
is not correct, as the two-particle threshold completely changes in
the Sarma phase, again due to the nonzero fermionic distribution functions
$f_{n\pm}>0$. As discussed in Appendix A, at $q\sim0$ the two-particle
threshold $\omega_{\textrm{2p}}^{(c)}$ should be about $2\delta\mu=\varepsilon_{F}$,
much larger than the naïve result of $2\left|\Delta_{c}\right|\simeq0.16\varepsilon_{F}$.
The large two-particle threshold, as shown in Fig. \ref{fig8}(k)
in cyan, ensures an undamped Leggett mode at low transferred momentum.

\section{Conclusions}

In summary, we have theoretically investigated the imbalanced superfluidity
of a three-dimensional strongly interacting Fermi gas near orbital
Feshbach resonances. The system can be well treated as a specific
realization of the two-band or two-channel model \cite{He2015TwoBandTheory},
with symmetric intra-channel inter-particle interactions. We have
found that by engineering the detuning (i.e., chemical potential)
of the closed channel via an external magnetic field, the induced
asymmetry in the single-particle dispersion relation between open
and closed channels can thermodynamically stabilize a Sarma pairing
in the closed channel. In three dimensions, we have predicted that
the resultant Sarma superfluid is robust against the formation of
a spatially inhomogeneous Fulde\textendash Ferrell\textendash Larkin\textendash Ovchinnikov
superfluid in the large spin-polarization limit. 

As a consequence of the gapless fermionic quasi-particle excitations,
the Sarma superfluid has a damped Goldstone-Anderson-Bogoliubov phonon
mode even at zero temperature. The damping rate of the phonon mode
becomes significant at moderate transferred momentum. The Sarma superfluid
also has a well-defined, undamped massive Leggett mode at low momentum,
due to the lifted two-particle continuum. However, as the transferred
momentum increases, the Leggett mode disappears once it enters the
particle-hole continuum. Experimentally, these peculiar features of
the collective modes of the Sarma superfluid can be measured by using
Bragg spectroscopy.

In one dimension or two dimensions, the Fulde\textendash Ferrell\textendash Larkin\textendash Ovchinnikov
superfluidity may become favorable due to the reduced dimensionality
\cite{Hu2007,Liu2007,Orso2007,Toniolo2017}. In that cases, we anticipate
a rich and complicated phase diagram \cite{Mizushima2013,Takahashi2014}.
Moreover, it is interesting to understand the pair fluctuations in
a strongly interacting Sarma superfluid, based on the standard Gaussian
pair fluctuation theory \cite{Hu2006EPL,Diener2008} or the functional
renormalization group \cite{Boettcher2015PLB,Strack2014,Boettcher2015PRA}.
These possibilities will be explored in future studies. 
\begin{acknowledgments}
Our research was supported by the National Natural Science Foundation
of China, Grant No. 11747059 (P. Z.) and Grant No. 11775123 (L. H.),
and by Australian Research Council's (ARC) Discovery Projects: FT140100003
and DP180102018 (X.-J. L), FT130100815 and DP170104008 (H. H.). L.
H. acknowledges the support of the Recruitment Program for Young Professionals
in China (i.e., the Thousand Young Talent Program).
\end{acknowledgments}

\appendix
%dummy comment inserted by tex2lyx to ensure that this paragraph is not empty

\section{The two-particle continuum of a Sarma superfluid}

In this appendix, we consider a Sarma superfluid with dispersion relations
$E_{\mathbf{k},+}=E_{\mathbf{k}}-\delta\mu$ (lower branch) and $E_{\mathbf{k},-}=E_{\mathbf{k}}+\delta\mu$
(upper branch), where $E_{\mathbf{k}}=\sqrt{\xi_{\mathbf{k}}^{2}+\Delta^{2}}$.
We aim to calculate the two-particle threshold of the Sarma superfluid,
\begin{equation}
\omega_{\textrm{2p}}\left(\mathbf{q}\right)=\min_{\{\mathbf{k}\}}\left[E_{\mathbf{k}}+E_{\mathbf{k}+\mathbf{q}}\right],
\end{equation}
under the condition that $E_{\mathbf{k}}\geq\delta\mu$ (see the second
and third terms in Eq. (\ref{eq: M11}) and Eq. (\ref{eq: M12}),
contributed from the two-particle excitations). At zero momentum $\mathbf{q}=0$,
we find immediately that $\omega_{\textrm{2p}}=2\delta\mu$. For nonzero
momentum $q$, let us assume that the single-particle dispersion relation
has a minimum at $k_{c}=\sqrt{2m\mu}/\hbar$ ($\mu>0$) and has a
zero at $k_{\textrm{ph}}=[2m(\mu+\sqrt{\delta\mu^{2}-\Delta^{2}})]^{1/2}/\hbar$
(see Fig. 9). Using $E_{\mathbf{k}+\mathbf{q}}=\sqrt{[\hbar^{2}(\mathbf{k}+\mathbf{q})^{2}/2m-\mu]^{2}+\Delta^{2}}$,
it is easy to obtain that,
\begin{equation}
\omega_{\textrm{2p}}=\left\{ \begin{array}{cc}
\delta\mu+\sqrt{\left[\frac{\hbar^{2}(k_{\textrm{ph}}-q)^{2}}{2m}-\mu\right]+\Delta^{2}}, & q<k_{\textrm{ph}}-k_{c}\\
\delta\mu+\left|\Delta\right|, & \left|q-k_{\textrm{ph}}\right|\leq+k_{c}\\
\delta\mu+\sqrt{\left[\frac{\hbar^{2}(k_{\textrm{ph}}-q)^{2}}{2m}-\mu\right]+\Delta^{2}}. & k_{\textrm{ph}}+k_{c}<q
\end{array}\right.\label{eq:w2p}
\end{equation}

\section{The particle-hole continuum of a Sarma superfluid}

We now turn to consider the particle-hole continuum. We want to determine,
\begin{equation}
\omega_{\textrm{ph}}\left(\mathbf{q}\right)=\max_{\{\mathbf{k}\}}\left[E_{\mathbf{k}+\mathbf{q}/2}-E_{\mathbf{k}-\mathbf{q}/2}\right],
\end{equation}
under the constraints $E_{\mathbf{k}+\mathbf{q}/2}>\delta\mu>E_{\mathbf{k}-\mathbf{q}/2}$
(see the first term in Eq. (\ref{eq: M11}) and Eq. (\ref{eq: M12})).
The collective mode couples to the particle-hole excitations and is
damped, if its frequency $0<\omega<\omega_{\textrm{ph}}$. These particle-hole
excitations occur at around $k_{\textrm{ph}}$. It is readily seen
that in order to reach the maximum, we must have $\mathbf{k}\parallel\mathbf{q}$
and $k=k_{\textrm{ph}}+q/2$. This leads to,
\begin{equation}
\omega_{\textrm{ph}}=\sqrt{\left[\frac{\hbar^{2}(k_{\textrm{ph}}+q)^{2}}{2m}-\mu\right]+\Delta^{2}}-\delta\mu.\label{eq:wph}
\end{equation}
It is worth noting that when $0<\omega<\omega_{\textrm{ph}}$ the
damping of the collective mode due to the particle-hole excitations
depends on the coherent factor $u_{\mathbf{k}\pm\mathbf{q}/2}v_{\mathbf{k}\pm\mathbf{q}/2}$
(see the first term in Eq. (\ref{eq: M11}) and Eq. (\ref{eq: M12})
for the matrix elements $M_{11}$ and $M_{12}$). The coherence factor
is significant at about $k_{c}$ only, as shown in Fig. 9, which implies
a resonant condition $k_{\textrm{ph}}-q/2\sim k_{c}$. This means
that the damping due to particle-hole excitations becomes important
at $q\sim2(k_{\textrm{ph}}-k_{c})$.

\end{document}